\documentclass[sigconf,nonacm]{acmart}
\AtBeginDocument{%
  }

\setcopyright{acmlicensed}
\copyrightyear{2018}
\acmYear{2018}
\acmDOI{XXXXXXX.XXXXXXX}
\acmConference[Conference acronym 'XX]{Make sure to enter the correct
  conference title from your rights confirmation email}{June 03--05,
  2018}{Woodstock, NY}
\acmISBN{978-1-4503-XXXX-X/2018/06}

\usepackage{xspace}

\usepackage{enumitem}

\usepackage{graphicx}
\usepackage{subcaption}

\usepackage{algorithm}
\usepackage{algorithmic}
\usepackage{amsmath}
\usepackage{xcolor}

\definecolor{algogreen}{RGB}{58, 130, 54}
\newcommand{\algcomment}[1]{\textcolor{algogreen}{\textit{// #1}}}

\newcommand{\SysName}{ForkKV\xspace}

\begin{document}

\title{\SysName: Scaling Multi-LoRA Agent Serving via Copy-on-Write Disaggregated KV Cache}

\author{Shao Wang}
\orcid{0009-0000-9816-2582}
\affiliation{%
  \institution{Shanghai Jiao Tong University}
  \city{Shanghai}
  \country{China}
}
\email{shaowang@sjtu.edu.cn}

\author{Rui Ren}
\affiliation{%
  \institution{Shanghai Jiao Tong University}
  \city{Shanghai}
  \country{China}
}
\email{renrui@sjtu.edu.cn}

\author{Lin Gui}
\affiliation{%
  \institution{Shanghai Jiao Tong University}
  \city{Shanghai}
  \country{China}
}
\email{guilin@sjtu.edu.cn}

\renewcommand{\shortauthors}{Shao et al.}

\begin{abstract}
The serving paradigm of large language models (LLMs) is rapidly shifting towards complex multi-agent workflows where specialized agents collaborate over massive shared contexts. While Low-Rank Adaptation (LoRA) enables the efficient co-hosting of these specialized agents on a single base model, it introduces a critical memory footprint bottleneck during serving. Specifically, unique LoRA activations cause Key-Value (KV) cache divergence across agents, rendering traditional prefix caching ineffective for shared contexts. This forces redundant KV cache maintenance, rapidly saturating GPU capacity and degrading throughput.

To address this challenge, we introduce \textsc{\SysName}, a serving system for multi-LoRA agent workflows centered around a novel memory management paradigm in OS: \textit{fork with copy-on-write} (CoW). By exploiting the structural properties of LoRA, \textsc{\SysName} physically decouples the KV cache into a massive shared component (analogous to the parent process's memory pages) and lightweight agent-specific components (the child process's pages). To support this mechanism, we propose a \textit{DualRadixTree} architecture that allows newly forked agents to inherit the massive shared cache and apply CoW semantics for their lightweight unique cache. Furthermore, to guarantee efficient execution, we design \textit{ResidualAttention}, a specialized kernel that reconstructs the disaggregated KV cache directly within on-chip SRAM. Comprehensive evaluations across diverse language models and practical datasets of different tasks demonstrate that \textsc{\SysName} achieves up to 3.0x the throughput of state-of-the-art multi-LoRA serving systems with a negligible impact on generation quality.

\end{abstract}




\maketitle
\section{Introduction}

The serving paradigm of Large Language Models (LLMs) has rapidly evolved from simple chatbots to complex autonomous agentic workflows, such as coding assistants equipped with advanced reasoning, planning, and tool-call capabilities~\cite{cursor,claudecode,codex}. These workflows operate as collaborative pipelines of specialized LLM nodes, termed \textit{agents}, each dedicated to different subtasks. A defining characteristic of these multi-agent workloads is their context structure. Agents typically share a massive static context, often dominated by a lengthy prefix like extensive system prompts or a large codebase~\cite{zheng2024sglang,yang2024sweagent,liu2023agentbench}. From this shared prefix, agents fork their own distinct context from dynamic outputs, such as reasoning steps from previous agents and historically acquired tool observations. For example, to resolve a Github issue, a coding assistant uses codebase as its shared static context and sequentially triggers navigation, generation, and testing agents, where each agent builds their own context based on previous agents. However, successfully executing such diverse subtasks requires distinct agent capabilities. A single monolithic model often lacks the flexibility to handle every stage optimally, necessitating fine-tuning the base model with task-specific datasets to serve specialized agents effectively~\cite{zeng2024agenttuning,chen2023fireact,schick2023toolformer,patil2024gorilla}.

To tailor foundational models for these diverse tasks in a workflow, Parameter-Efficient Fine-Tuning (PEFT)~\cite{peft} techniques, particularly Low-Rank Adaptation (LoRA)~\cite{hu2022lora}, offer a promising solution. By freezing the pretrained weights and updating only small low-rank matrices known as \textit{adapters}, LoRA maintains high generation quality while introducing minimal parameter overhead~\cite{hu2022lora,dettmers2023qlora,liu2024dora}. For example, LoRA adapters with a low-rank dimension of 16 trained on Llama3.1-70B~\cite{meta2024llama31} account for only 0.28\% of the original model size (approximately 400MB v.s. 140GB). Such extreme memory efficiency fundamentally shifts how systems manage concurrent workloads. Instead of deploying multiple monolithic models, the serving engine can concurrently host multiple specialized agents on a single shared base model by dynamically swapping these lightweight adapters. Many modern agentic applications have already adopted this efficient architectural design~\cite{liu2024droidspeak,yu2024neeko,liu2025videomind,kadekodi2025agentflux}, which we refer to as \textit{multi-LoRA agent serving} in this paper.

\begin{figure}[tbp]
  \centering
  \includegraphics[width=0.4\textwidth]{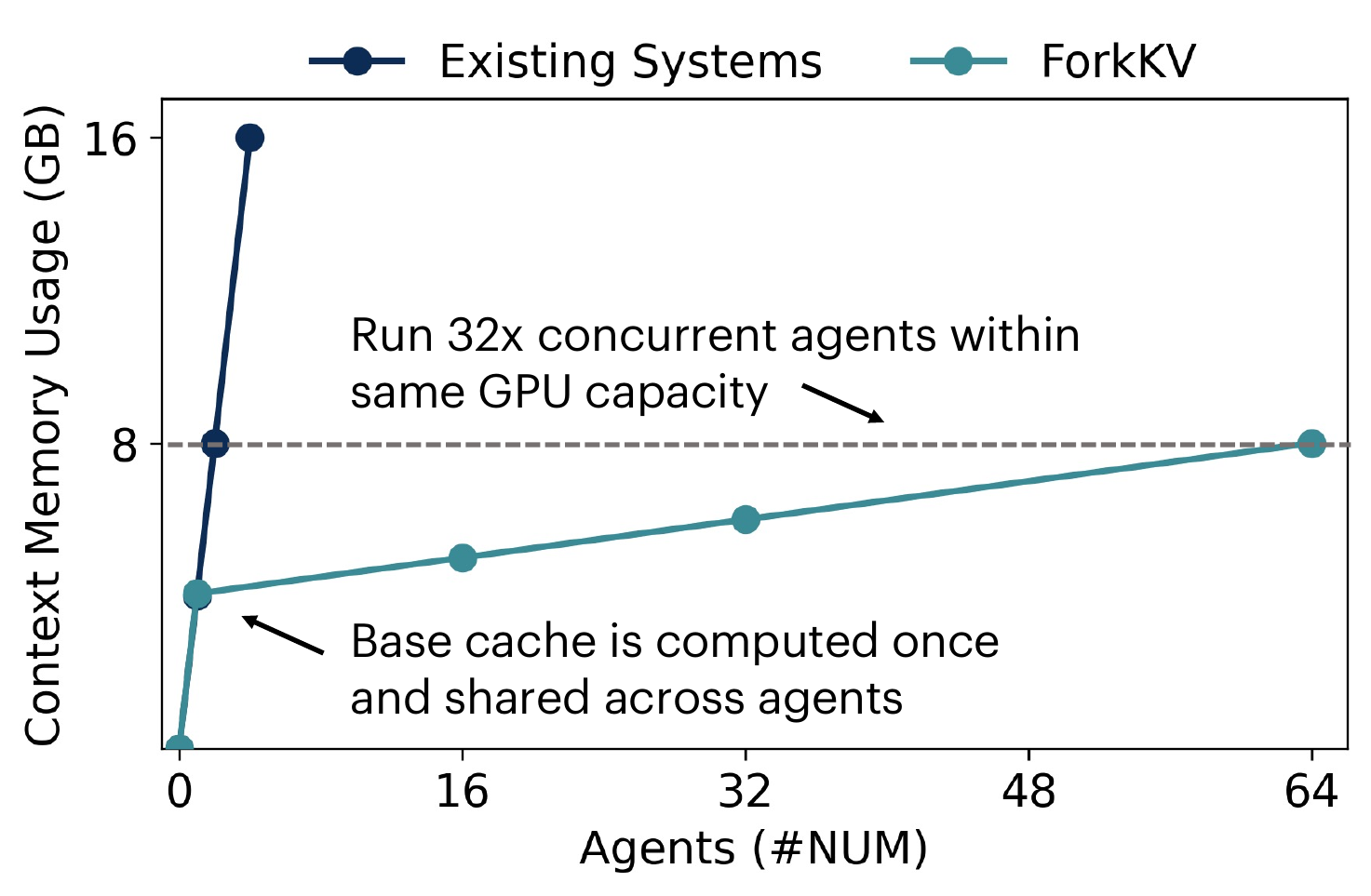}
  \caption{Context memory usage when serving agents with 32K shared contexts on Llama3-8B, where agents are based on different LoRA adapters and the rank is 16.}
  \label{fig:memory-comparsion}
\end{figure}

While LoRA successfully minimizes the memory overhead of model weights, multi-agent serving still suffers from a critical memory footprint bottleneck caused by the duplicated unshareable Key-Value states (i.e., KV cache) across agents. In traditional monolithic deployment, serving engines avoid memory redundancy through \textit{prefix caching}~\cite{zheng2024sglang,google2025prefixcache,openai2025prefixcache,anthropic2025prefixcache}. This technique allows different requests to share the KV cache of common prefixes. However, this optimization breaks down in multi-LoRA scenarios. The unique activations produced by each adapter cause the KV cache to diverge across different agents. Consequently, the system is forced to maintain an independent KV cache for each agent even when they process the exact same context, incurring severe memory footprint overhead. As shown in Figure~\ref{fig:memory-comparsion}, the overall memory consumption (represented by the deep blue line) scales linearly with the number of agents, rapidly saturating GPU capacity. Specifically, in sequential workflows like ReAct~\cite{yao2022react}, context reuse fails entirely when the pipeline switches to a new LoRA agent. In parallel workflows like MapReduce~\cite{luo2025autellix}, broadcasting the shared input creates redundant cache copies. Complex agent workflows exacerbate this redundancy, which could lead to severe performance degradation in both latency and throughput~\cite{yao2025cacheblend,wu2024loongserve,zhong2024distserve,agrawal2023sarathi}.

To overcome this memory footprint bottleneck, our key insight is that we can partially share the KV cache across agents by physically decoupling it based on the structural properties of LoRA. We term this partially shared memory layout \textit{disaggregated KV cache}. The standard LoRA projection $xW + xA_iB_i$ intuitively suggests separating the KV cache into a massive base cache ($bCache = xW$) and a lightweight adapter-specific residual cache ($rCache = xA_i$) that can reconstruct full projection via $rCache \times B_i$. This decomposition exposes a significant size asymmetry, where \textit{bCache} is typically dozens of times the size of \textit{rCache} due to the low-rank projection of LoRA. Thus, instead of redundantly allocating complete KV cache for agents processing the same context, we globally share the massive $bCache$ across agents and only maintain lightweight \textit{rCache} for each agent. This disaggregated design drastically cuts the per-agent memory footprint, effectively mitigating the memory footprint bottleneck. As validated in Figure~\ref{fig:memory-comparsion}, our approach enables an 8GB KV cache to support 32$\times$ more concurrent agents.

Admittedly, sharing $bCache$ beyond the first layer is mathematically lossy because adapter-specific activations cause subsequent inputs $x$ to diverge across agents. However, this divergence is empirically bounded. The transformer architecture provides inherent robustness through its residual connections~\cite{elhage2021mathematical}, and the LoRA adapters preserve necessary task-specific interactions~\cite{hu2022lora,wang2023orthogonal}. Consequently, this aggressive memory optimization maintains a similarity for input $x$ of over 99.4\%, yielding a negligible degradation in generation quality of only 1.60\% (see details in Figure~\ref{fig:motivation-accuracy-loss}).

Supporting disaggregated KV cache presents several major challenges. First, the system needs to manage the distinct lifecycles of the shared \textit{bCache} and unique \textit{rCache} while maintaining the structural dependencies imposed by their mathematical decomposition. As multi-agent collaboration naturally forms multi-branch reasoning paths, tracking the resulting 1-to-N base-to-residual mappings introduces severe complexity. Second, computing attention scores requires reconstructing KV cache from its disaggregated components, but naive reconstruction in HBM incurs severe memory and computational overhead. Materializing the full-sized KV cache in HBM for every agent will completely negate the memory savings. Conversely, performing in-place updates on the shared \textit{bCache} using \textit{rCache} causes memory access conflicts across agents, forcing sequential execution and destroying intra-batch parallelism.

To address the first management challenge, we introduce \textsc{\SysName}, a multi-LoRA agent serving system featuring a novel disaggregated KV cache management mechanism inspired by the operating system (OS) primitive for subprocess creation: \textit{fork with copy-on-write (CoW)}. \textsc{\SysName} manages the disaggregated KV cache with the same efficiency as an OS fork: the massive \textit{bCache} acts as the shareable and read-only memory pages of a parent process, the lightweight \textit{rCache} serves as the unique CoW footprint of a child process, and their dynamic combination represents the complete memory space. To orchestrate this dual-tiered layout in practice, \textsc{\SysName} introduces a coordinated \textit{DualRadixTree} architecture. When a new agent is launched, \textsc{\SysName} performs a longest-prefix match to inherit the globally shared read-only \textit{bCache}, forks the memory space by allocating memory exclusively for the agent's unique \textit{rCache}, and updates this dual-tree storage after generation.

To tackle the second challenge, we propose fusing KV cache reconstruction directly into the attention kernel. By keeping all intermediate computations within the fast on-chip SRAM, the design eliminates both the extra HBM allocation and the serialized execution caused by conflicting memory access. Based on this insight, we implement \textit{ResidualAttention}, an attention kernel specifically tailored for the disaggregated KV cache architecture. In the first step, the kernel streams \textit{bCache} and \textit{rCache} directly into SRAM in a block-wise manner and reconstructs Key cache. In the second step, \textit{ResidualAttention} computes separate attention scores for the base and residual components and fuses the final output by leveraging matrix associativity. 

In this work, we implement \SysName on top of SGLang~\cite{zheng2024sglang}, a state-of-the-art LLM serving framework for production. To assess the system, we design end-to-end evaluations based on two typical agentic serving scenarios, ReAct and MapReduce. We conduct these experiments across a wide range of LLMs, including Llama3-8B~\cite{llama3modelcard}, Qwen2.5-7B~\cite{qwen2.5}, and Qwen2.5-14B~\cite{qwen2.5}. Across practical workloads such as LooGLE~\cite{li2023loogle}, NarrativeQA~\cite{kocisky2018narrativeqa}, and APIGen~\cite{liu2024apigen}, \textsc{\SysName} demonstrates significant performance gains over state-of-the-art multi-LoRA serving systems. Specifically, \textsc{\SysName} achieves 1.25-3.04$\times$ the throughput on ReAct workflows, and 1.68-2.60$\times$ the throughput on MapReduce workflows, with a negligible quality degradation of only 0.71\% on average measured by F1-Score~\cite{ffl2020evalqa}.

In summary, we make the following contributions:

\begin{itemize}[topsep=3pt]
    \item We identify the memory footprint bottleneck in multi-LoRA agent serving, where adapter-specific KV cache divergence makes prefix caching ineffective.
    \item We propose \SysName, a multi-LoRA agent serving system inspired by the OS fork primitive with copy-on-write, utilizing a DualRadixTree to disaggregate the KV cache into a shareable base and LoRA-specific residuals.
    \item We design \textit{ResidualAttention}, an attention kernel fusing KV cache reconstruction for disaggregated KV cache layout.
    \item We comprehensively evaluate \SysName across diverse LLMs and datasets, demonstrating significant improvement over state-of-the-art LoRA serving systems.
\end{itemize}

\section{Background}

We first introduce the mechanics of LLM serving (\S\ref{sec:bg-serving}), then analyze the structural properties of LoRA (\S\ref{sec:bg-lora}), and finally discuss the computational demands of modern agentic workflows (\S\ref{sec:bg-agent}).

\subsection{LLM Serving}
\label{sec:bg-serving}

LLMs~\cite{llama3modelcard,qwen2.5,qwen3,deepseekv3,deepseekr1} predominantly adopt the Transformer architecture to generate text auto-regressively. During generation, tokens interact with historical context via attention mechanism~\cite{vaswani2017attention,ainslie2023gqa,shazeer2019fast}, where sequential order is typically captured by applying Rotary Position Embedding (RoPE)~\cite{su2024roformer} to the Query ($Q$) and Key ($K$) representations.

To avoid redundantly recomputing $K$ and $V$ tensors for historical tokens at every step, inference engines employ a \textit{KV cache}. This optimization naturally divides the serving process into two phases: a compute-heavy \textit{prefill phase} that processes the prompt to populate the initial KV cache, and a memory-bound \textit{decode phase} that auto-regressively generates new tokens by attending to the cached history. This efficiency is further extended across different requests via \textit{prefix caching}~\cite{zheng2024sglang,google2025prefixcache,openai2025prefixcache,anthropic2025prefixcache}. By identifying and reusing the KV cache of shared text segments (e.g., system prompts or shared context), inference engines significantly accelerate the Time-to-First-Token (TTFT) and optimize overall memory usage~\cite{yao2025cacheblend,wu2024loongserve,zhong2024distserve,agrawal2023sarathi}.

\subsection{Low-Rank Adaptation (LoRA)}
\label{sec:bg-lora}

Low-Rank Adaptation (LoRA)~\cite{hu2022lora} is a prominent Parameter-Efficient Fine-Tuning (PEFT)~\cite{peft} designed to mitigate the significant computational and memory costs associated with full-parameter fine-tuning. Instead of updating the entire model, LoRA freezes the pretrained weights and injects small trainable low-rank adapter matrices into transformer layers.

Formally, for a pretrained weight matrix $W \in \mathbb{R}^{m \times n}$, LoRA $i$ introduces two low-rank matrices $A_i \in \mathbb{R}^{m\times r}$ and $B_i \in \mathbb{R}^{r\times n}$, where the rank $r \ll m,n$. The combined projection is computed as:

\begin{align}
    Y &= xW+xA_iB_i
\end{align}

\noindent where $x \in \mathbb{R}^{s \times m}$ is the input hidden state, and $s$ is the number of tokens in a batch (batch size multiplied by sequence length).

This computation can be naturally decomposed into two distinct parts. We refer to the projection from the frozen weights, $xW \in \mathbb{R}^{s \times n}$, as the base model cache (\textit{bCache}), and the intermediate projection from the first low-rank matrix, $x A_i \in \mathbb{R}^{s \times r}$, as the residual cache (\textit{rCache}). The final projected state can thus be reconstructed via:

\begin{equation}
    Y = bCache + rCache \times B_i
\end{equation}

This algebraic decomposition reveals two critical properties. First, because $r \ll n$, the bCache is significantly larger than the rCache (e.g., 64 times larger given a typical $n=1024$ and $r=16$). Second, the output dimension ($r$) of the rCache inherently mismatches the dimension ($n$) required by the Rotary Position Embedding (RoPE) matrix $R_p$. Consequently, RoPE cannot be directly applied to the rCache. Its application must be deferred until the rCache is projected back to the full $n$-dimensional space via $B_i$.

\subsection{Agentic Workflow}
\label{sec:bg-agent}

Modern LLM applications have evolved into \textit{agentic workflows}, which are comprehensive pipelines that decompose complex problems into actionable steps. Each step is typically executed by an individual autonomous \textit{agent} node. A defining characteristic of these workflows, whether in sequential (e.g.,ReAct~\cite{yao2022react}) or parallel (e.g., MapReduce~\cite{luo2025autellix}), is their forked context structure: agents rely on a massive shared static prefix (e.g., a large codebase) and subsequently branch off to build distinct contexts from dynamic outputs like prior reasoning steps and tool observations.

As workflows grow more complex, the individual agents within them require highly specialized expertise. Because deploying a separate fully fine-tuned model for each specialized agent node is memory prohibitive~\cite{chen2024punica,sheng2023slora}, multi-LoRA serving has emerged as the prevailing paradigm~\cite{liu2024droidspeak,yu2024neeko,liu2025videomind,kadekodi2025agentflux}. By multiplexing task-specific LoRA modules on a single shared base model, systems can efficiently support the diverse agents constituting the workflow while minimizing memory overhead.

\section{Motivation}
\label{sec:motivation}

In this section, we first highlight the critical memory footprint bottlenecks of prefix caching in multi-LoRA serving (\S\ref{sec:motivation-prefix}). We then demonstrate the potential of context sharing via disaggregated KV cache (\S\ref{sec:motivation-sharing}), followed by an analysis of the two primary system challenges this approach introduces (\S\ref{sec:motivation-challenges}): the management of disaggregated KV cache, and the memory footprint and computational overhead caused by KV cache reconstruction.

\subsection{Inefficient Prefix Caching}
\label{sec:motivation-prefix}

Prefix caching significantly reduces Time-to-First-Token (TTFT) and improves throughput in modern LLM serving systems~\cite{yao2025cacheblend,wu2024loongserve,zhong2024distserve,agrawal2023sarathi}. However, this mechanism relies on the assumption that identical text prefixes yield identical KV cache. This assumption fails in multi-LoRA serving scenarios where requests target distinct adapters. Unique LoRA activations cause KV cache divergence even if the text prefixes are exactly the same. Consequently, the generated KV cache becomes strictly tied to specific adapters and can no longer be shared. The system is therefore forced to compute and store duplicated KV cache for each adapter, rendering traditional prefix caching ineffective.

\begin{figure}[t]
    \centering
    \includegraphics[width=0.8\linewidth]{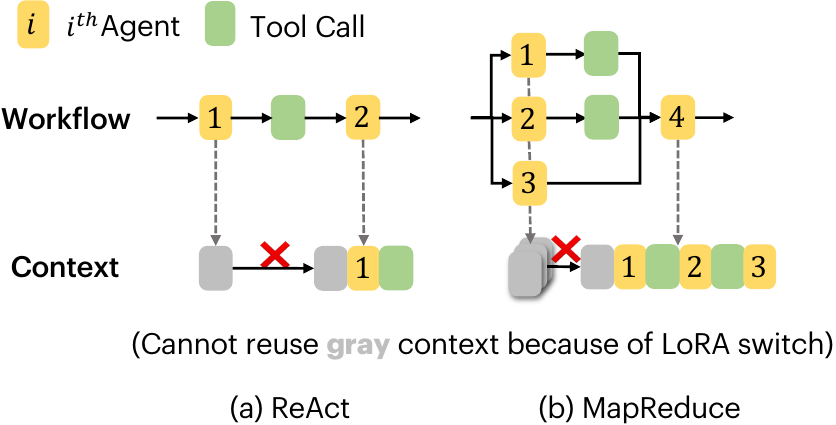}
    \caption{Failure of reusing KV cache across individual agents using different LoRA adapters.}
    \label{fig:motivation-context-reuse}
\end{figure}

\begin{figure}[t]
    \centering
    \includegraphics[width=0.6\linewidth]{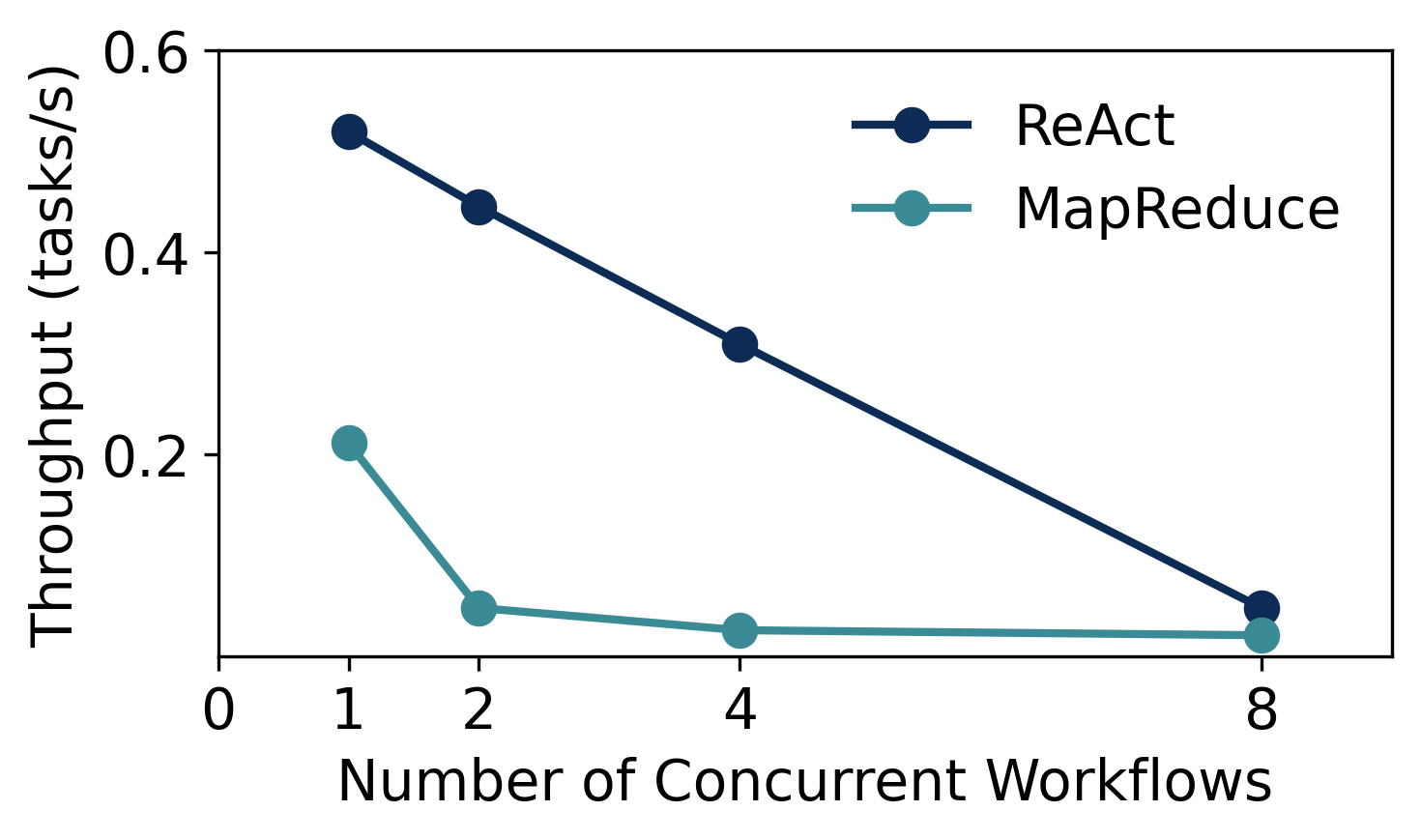}
    \caption{End-to-end throughput of prefix caching with different number of concurrent workflows. Every individual agent in different workflows uses different LoRA adapters.}
    \label{fig:motivation_lora}
\end{figure}

This redundancy becomes exceptionally severe in multi-LoRA agent serving scenarios. The execution context in these workflows is typically dominated by massive shared static inputs such as system prompts, codebases, or internal documents~\cite{liu2024droidspeak}, alongside continuously appended intermediate steps. Figure \ref{fig:motivation_lora} illustrates the throughput of ReAct and MapReduce workflows operating on 32K contexts across varying numbers of distinct concurrent workflows, where each workflow utilizes a completely non-overlapping set of LoRA adapters. As the number of workflows scales from 1 to 8, the throughput for ReAct and MapReduce drops by 90.8\% and 90.1\% respectively. This degradation occurs because, in both sequential workflows like ReAct (Figure \ref{fig:motivation-context-reuse}a) and parallel workflows like MapReduce (Figure \ref{fig:motivation-context-reuse}b), the inability to share KV cache across distinct LoRA adapters means agents must repeatedly process the massive static inputs alongside incremental contexts and maintain independent KV cache at each reasoning step, which not only inflates TTFT but also drastically increases memory consumption. As concurrency increases, these redundant KV cache rapidly exhaust GPU memory. This high memory pressure leaves little space for other active requests, severely restricting batch parallelism and causing the observed throughput collapse.

\vspace{1em}
\noindent\textbf{Takeaway \#1: }\textit{Prefix caching is inefficient in multi-LoRA agent serving scenarios because KV cache cannot be shared across agents with different LoRA adapters, resulting in critical throughput degradation.}

\subsection{Opportunities of Sharing Contexts}
\label{sec:motivation-sharing}

To mitigate memory contention in multi-LoRA agent serving, our key insight is to replace the traditional unified KV cache with a physically decoupled architecture. We introduce a partially shared memory layout termed the \textit{disaggregated KV cache}. As established in Section \ref{sec:bg-lora}, the standard LoRA projection $xW + xA_iB_i$ naturally decomposes the attention states into a massive base cache ($bCache = xW$) and a lightweight residual cache ($rCache = xA_i$) that can reconstruct the LoRA projection through $rCache \times B_i$. This decomposition exposes a significant size asymmetry, where \textit{bCache} is typically dozens of times the size of \textit{rCache} due to the low-rank projection of LoRA.

\begin{figure}
    \centering
    \includegraphics[width=0.98\linewidth]{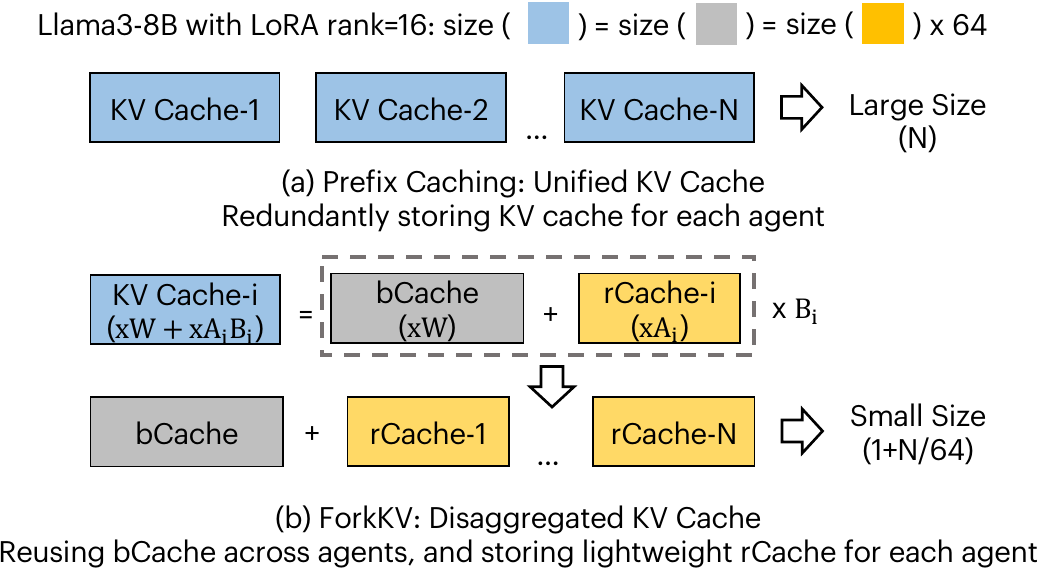}
    \caption{Memory usage comparison of \textsc{ForkKV} v.s. prefix caching when serving N agents with distinct LoRAs (r=16) on Llama3-8B. \textit{bCache} and \textit{rCache} represent base cache and residual cache respectively in disaggregated KV cache layout.}
    \label{fig:motivation-comparison}
\end{figure}

Our design directly exploits this structural property. Traditional prefix caching, as shown in Figure~\ref{fig:motivation-comparison}a, redundantly allocates a complete KV cache for each concurrent agent. Instead, our approach computes the large $bCache$ exactly once and shares it globally across all agents processing identical contexts. Each individual agent then only allocates memory for its own small $rCache$. This theoretical advantage translates into concrete system benefits. Consider 16 concurrent agents processing the same 32K context on Llama3-8B. Traditional methods require 4GB of GPU memory per agent, consuming 64GB in total. In contrast, the disaggregated layout requires only a single 4GB $bCache$ alongside 16 unique 64MB $rCache$ allocations. This reduces the total memory consumption to approximately 5GB, yielding an 11.8$\times$ memory efficiency improvement. Consequently, the disaggregated architecture transforms the inherent decomposability of the KV cache into a critical optimization opportunity, fundamentally alleviating memory contention in multi-LoRA agent serving.

\begin{figure}[t]
    \centering
    \begin{subfigure}{0.48\linewidth}
        \centering
        \includegraphics[width=\linewidth]{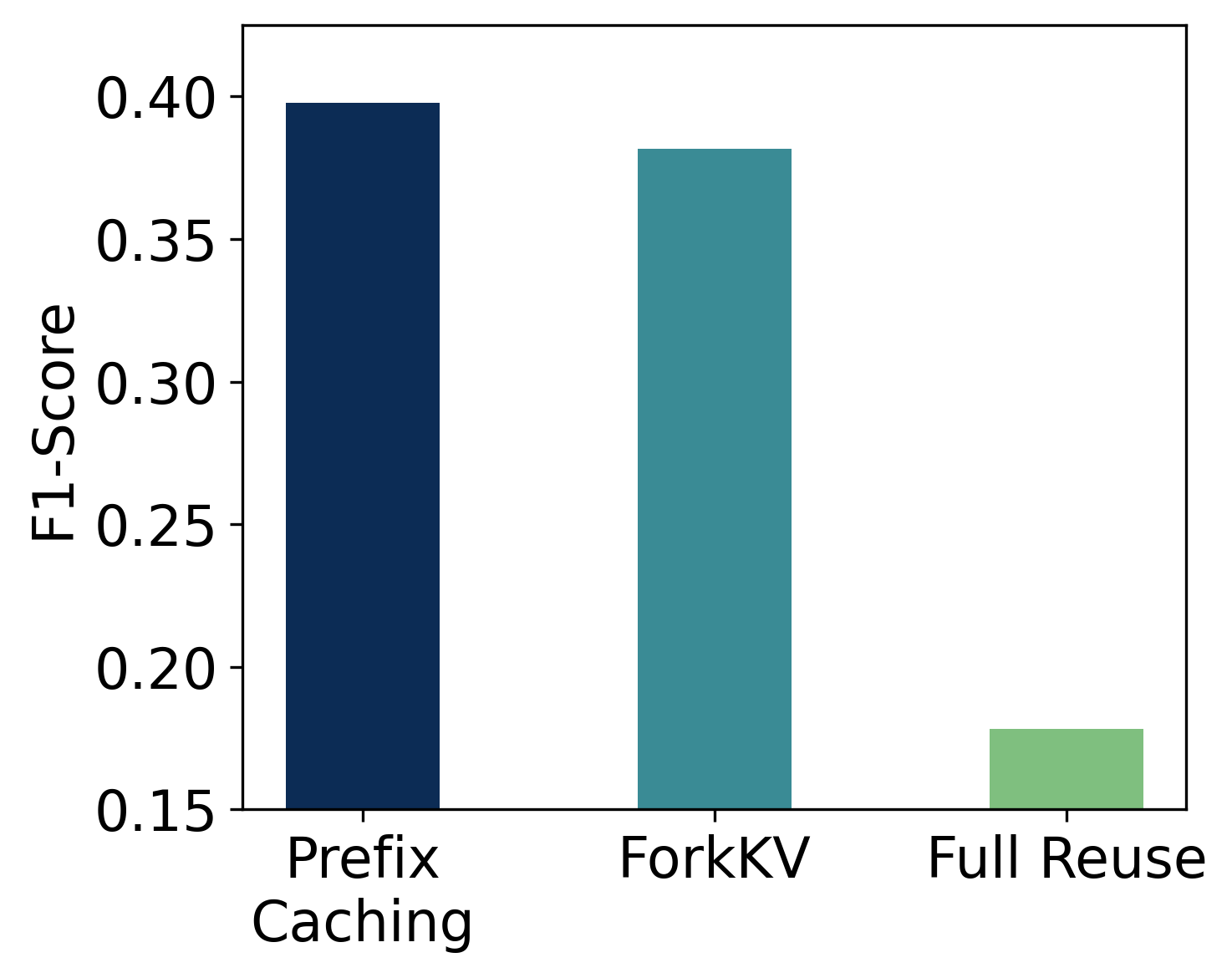}
        \caption{Generation Quality}
        \label{fig:motivation-quality}
    \end{subfigure}
    \hfill
    \begin{subfigure}{0.48\linewidth}
        \centering
        \includegraphics[width=\linewidth]{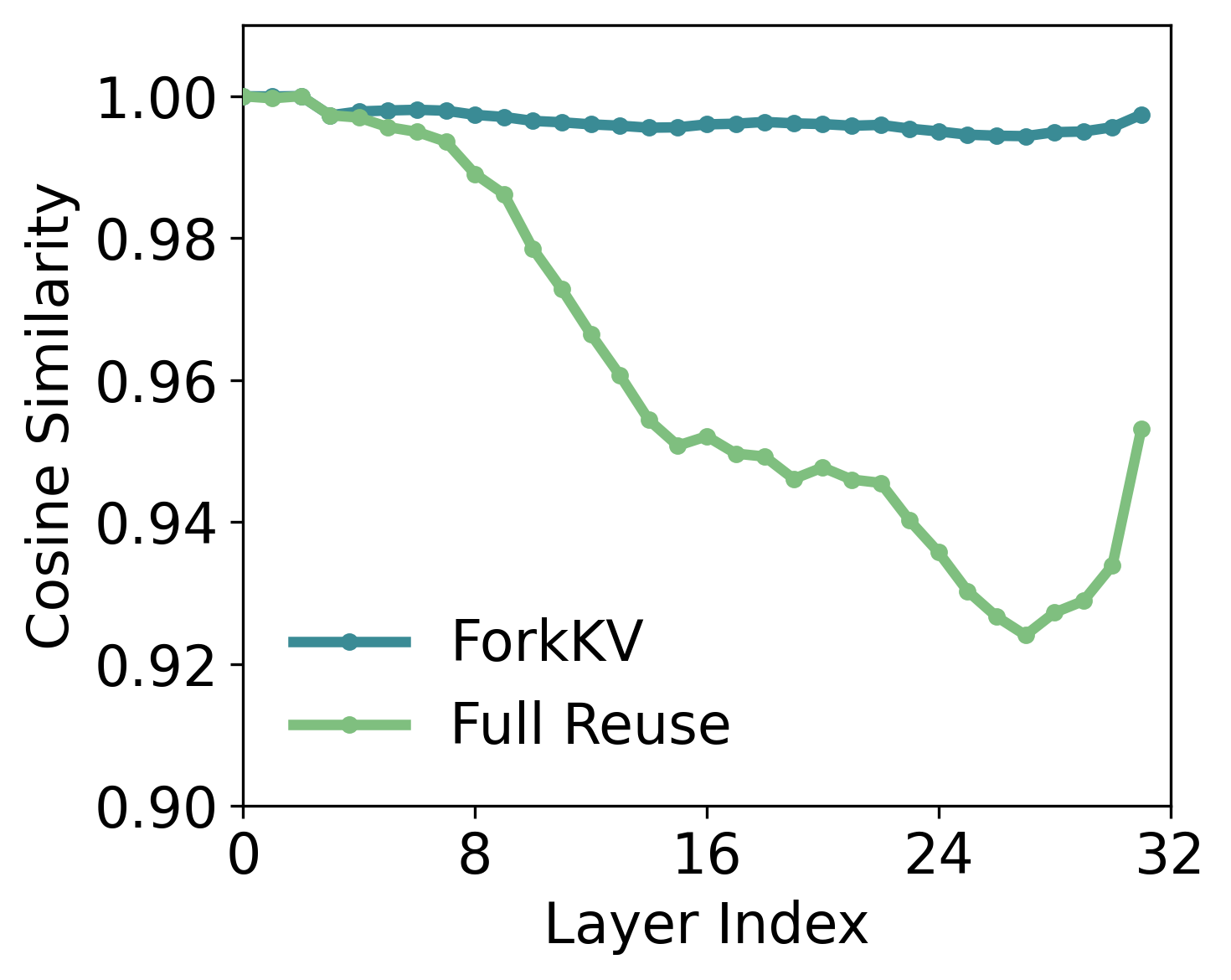}
        \caption{Similarity of Input x}
        \label{fig:motivation-similarity}
    \end{subfigure}
    \caption{\textsc{ForkKV} maintains comparable (a) generation quality and (b) input x similarity compared to prefix caching. Experiments are conducted on \textit{APIGen} dataset.}
    \label{fig:motivation-accuracy-loss}
\end{figure}

\noindent \textbf{Negligible Accuracy Loss.} While the disaggregated KV cache significantly improves memory efficiency, sharing a unified base cache beyond the first layer is mathematically lossy. This occurs because adapter-specific activations at each transformer layer cause the exact input state $x$ to diverge across agents. However, this approximation introduces minimal error in practice, yielding only a 1.60\% accuracy loss on tool calling benchmarks like \textit{APIGen}, as demonstrated in Figure~\ref{fig:motivation-quality}. This minimal impact is driven by two key factors. First, the residual connections $x_{l+1}=x_l+F(x_l)$ within the transformer architecture prevent the base state $x_l$ from radical drift~\cite{elhage2021mathematical}. Second, our disaggregated layout explicitly preserves the task-specific interactions between Q, K and V within each LoRA adapter, which are necessary for accurate attention outputs. While residual connections bound the overall state divergence, high generation quality still requires the attention mechanism to correctly extract task-specific features. Previous studies demonstrate that the effectiveness of LoRA relies on joint optimization of these QKV projections~\cite{hu2022lora,wang2023orthogonal}. Our disaggregated layout explicitly preserves this mechanism by computing a dedicated adapter cache for each agent. Therefore, every adapter applies its unique K and V transformations to the shared context. This guarantees accurate attention outputs and prevents errors from compounding in the residual stream. As a result, our method maintains an input state cosine similarity of over 99.4\% across all layers compared to standard prefix caching (Figure~\ref{fig:motivation-similarity}). In contrast, the full reuse baseline entirely shares the KV cache and breaks the necessary QKV coadaptation. This leads to inaccurate attention outputs that progressively accumulate, dropping the cache similarity to approximately 92.4\% and causing a severe 21.0\% accuracy loss (Figure~\ref{fig:motivation-accuracy-loss}).

\noindent \textbf{Limitations of Prior Approaches.} Existing multi-LoRA agent serving systems are fundamentally trapped in a dilemma between memory scalability and generation accuracy. Maintaining independent KV cache for each adapter preserves accuracy but incurs prohibitive memory overhead, making high-concurrency deployments unfeasible. Conversely, as illustrated above, entirely sharing the cache through a full reuse paradigm alleviates memory pressure but severely degrades generation quality. To navigate this bottleneck, recent studies have explored selective KV cache recomputation~\cite{yao2025cacheblend,gim2024promptcache}. However, these techniques are explicitly designed for single-model environments and fail to manage the state divergence across multiple agents. While DroidSpeak~\cite{liu2024droidspeak} extends cache sharing to multi-model scenarios by recomputing critical layers, it still treats the KV cache as an indivisible monolithic unit, missing a critical opportunity to minimize the memory footprint through structural decoupling. Overall, these fundamental limitations necessitate a new disaggregated KV cache architecture tailored for multi-LoRA agent serving.

\vspace{1em}
\noindent\textbf{Takeaway \#2: }\textit{Replacing the monolithic KV cache with a shared base and an isolated residual significantly reduces memory consumption for concurrent agents without compromising on generation quality.}

\subsection{Challenges}
\label{sec:motivation-challenges}

While a disaggregated KV cache architecture maximizes context sharing with negligible accuracy loss, realizing these benefits exposes two fundamental challenges in system design.

\noindent \textbf{Managing Disaggregated KV Cache.} The primary challenge lies in managing the distinct lifecycles of the \textit{bCache} and \textit{rCache} while maintaining the structural dependencies imposed by their mathematical decomposition. This difficulty stems from their different access patterns. The massive \textit{bCache} acts as a shared foundation accessed by multiple concurrent agents. Conversely, the lightweight \textit{rCache} is strictly tied to individual agents using different LoRA adapters. Existing serving systems~\cite{kwon2023efficient,zheng2024sglang,sheng2023slora} typically assume an indivisible KV cache. They manage these states under a single lifecycle within a unified memory pool, inherently lacking the structural abstraction required for a disaggregated memory architecture. Furthermore, as multi-agent collaboration naturally forms multi-branch reasoning paths, tracking the resulting 1-to-N base-to-residual mappings within a unified pool introduces severe complexity when resolving context dependencies. These structural limitations therefore necessitate a novel caching abstraction specifically tailored for the disaggregated KV cache layout.

\noindent \textbf{KV Cache Reconstruction.} Although decomposing the KV cache improves memory efficiency, computing accurate attention scores requires reconstructing KV cache from its disaggregated components. This requirement introduces a critical challenge because naive HBM-based reconstruction methods incur severe memory footprint and computational overhead. For instance, materializing a full-sized KV cache in HBM for every agent prior to attention computation completely negates our intended memory savings. An alternative approach involves directly updating the shared \textit{bCache} in place using LoRA projections recovered from the \textit{rCache}. However, this operation causes memory access conflicts across concurrent agents, forcing sequential execution and destroying intra-batch parallelism. Consequently, we must design a reconstruction mechanism that preserves both memory efficiency and batched execution, thereby enabling high-performance inference under the disaggregated KV cache architecture.

\section{Overview and Key Ideas}

\begin{figure}
    \centering
    \includegraphics[width=0.85\linewidth]{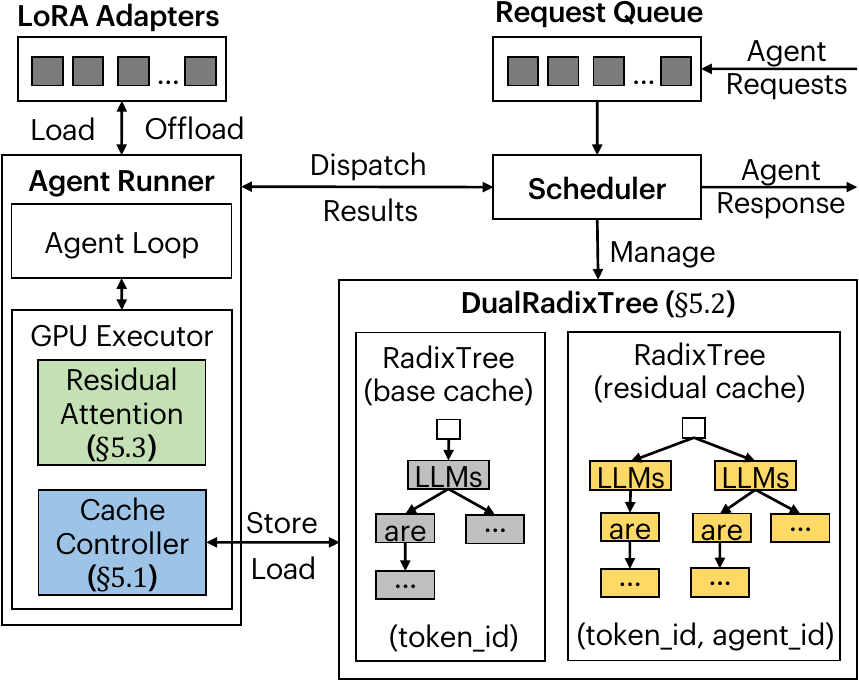}
    \caption{Overview of \textsc{\SysName}}
    \label{fig:overview}
\end{figure}

To fully exploit the opportunities of context sharing in multi-LoRA agent serving scenarios, we need to address two challenges: the management of disaggregated KV cache, and the memory footprint and computational overhead caused by KV cache reconstruction. In this paper, we propose \textsc{\SysName}, a multi-LoRA agent serving system that improves throughput using an OS-inspired disaggregated cache management mechanism and a novel attention kernel. In this section, we summarize two key ideas behind our system:

\noindent \textbf{1. OS-Inspired Disaggregated Cache Management.} Managing a disaggregated KV cache requires a novel memory abstraction. Traditional unified pools fail to accommodate the distinct lifecycles and structural dependencies of shared and unique memory components. To bridge this gap, \textsc{\SysName} introduces a DualRadixTree architecture. As shown in Figure~\ref{fig:overview}, this structure natively supports the disaggregated layout by physically separating the management of the massive base cache from the lightweight residual cache.

\begin{figure}[t]
    \centering
    \begin{subfigure}{0.47\linewidth}
        \centering
        \includegraphics[width=\linewidth]{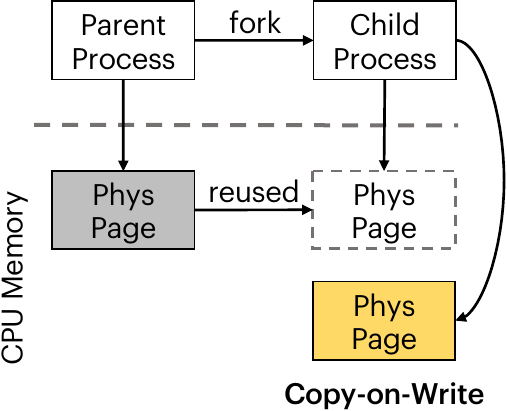}
        \caption{OS: Create child process}
        \label{fig:design-os-cow}
    \end{subfigure}
    \hfill
    \begin{subfigure}{0.49\linewidth}
        \centering
        \includegraphics[width=\linewidth]{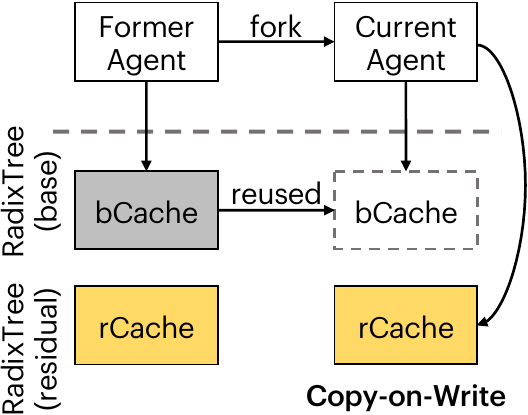}
        \caption{\SysName: Create new agent}
        \label{fig:design-forkv-cow}
    \end{subfigure}
    \caption{\textsc{\SysName} use OS-inspired fork semantics to create memory space for new agents with copy-on-write.}
    \label{fig:design-cow}
\end{figure}

To orchestrate this decoupled memory architecture, we introduce an operating system inspired fork semantics with copy-on-write. As demonstrated in Figure \ref{fig:design-cow}, when a new agent is launched, it forks the cache state from an existing agent in two steps. First, the new agent inherits the globally shared base cache which contains the same context in the base radix tree. This process is analogous to a newly forked OS process mapping the read-only physical pages of its parent. Second, the agent allocates exclusive memory for its unique residual cache in the residual radix tree. This part serves as the child process's isolated copy-on-write footprint for the new agent. Together, the dual-tree design and fork semantics translate our theoretical decoupling into a practical system design that systematically resolves the memory management complexity of disaggregated KV cache.

\noindent \textbf{2. ResidualAttention.} KV cache reconstruction incurs severe memory and computational overhead when operating in HBM. Previous naive methods struggle with either a prohibitive HBM allocation or a severe degradation of intra-batch parallelism. To address this challenge, we propose fusing KV cache reconstruction directly into the attention kernel. By keeping all intermediate computations within the fast on-chip SRAM, this design eliminates both the extra HBM allocation and the serialized execution caused by conflicting memory accesses.

Based on this insight, we implement \textit{ResidualAttention}, an attention kernel specifically tailored for the disaggregated KV cache architecture. Rather than assuming a monolithic KV cache layout, ResidualAttention streams the decoupled \textit{bCache} and \textit{rCache} directly into the SRAM in a block-wise manner. Inside the SRAM, the kernel first reconstructs the LoRA residuals of K using up-projection and a deferred RoPE operation. It then computes attention scores separately for the base and residual components and fuses the final output by leveraging matrix associativity. Through this algorithm and system co-design, ResidualAttention preserves the memory savings enabled by the disaggregated KV cache while ensuring high-throughput batched execution for concurrent agents.

\noindent \textbf{System Workflow.} Figure~\ref{fig:overview} illustrates the execution pipeline of \textsc{\SysName}. Upon receiving an agent request, the system enqueues it into the queue. The scheduler retrieves the request, parses the agent context, and then queries the DualRadixTree to perform a prefix matching against the existing base cache. To construct the memory state for the new agent, the system inherits this matched base cache and allocates an exclusive memory region for the residual cache of the specified adapter. Following memory allocation, the scheduler dispatches the request to the agent runner, where \textsc{\SysName} loads the requested LoRA adapters and establishes the agent loop to interleave model reasoning with external tool invocations. At the core of the agent runner is the GPU executor, which serves concurrent inference workloads across multiple agents. During execution, the cache controller directly loads and stores the base and residual cache according to the memory regions assigned by the scheduler. Operating on these retrieved data blocks, our custom \textit{ResidualAttention} kernel computes attention natively over the disaggregated memory layout. Finally, the agent runner returns the generated outputs to the scheduler to formulate the client response.

\section{System Design}
\label{sec:design}

\subsection{Disaggregating KV Cache}
\label{sec:design-kv-cache}

\begin{figure}[t]
    \centering
    \includegraphics[width=0.75\linewidth]{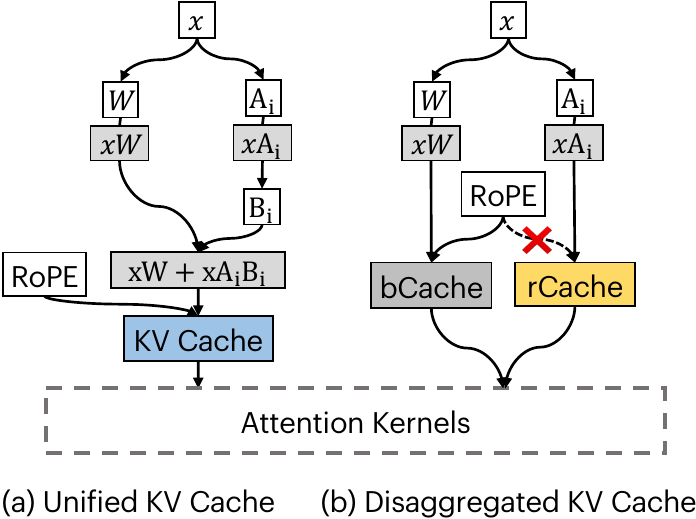}
    \caption{Unified v.s. Disaggregated KV Cache. Note that for simplicity, the cache blocks represent the storage of both K and V, but RoPE is only applied to K prior to caching.}
    \label{fig:design-kv-cache}
\end{figure}

Current unified KV cache design introduces a severe memory footprint bottleneck during concurrent multi-LoRA agent serving. As illustrated in Figure~\ref{fig:design-kv-cache}(a), to generate Key and Value tensors, the system projects the input $x$ through both the base weight $W$ and the LoRA matrices $A_i$ and $B_i$. Crucially, for Key tensors, the Rotary Position Embedding (RoPE) is applied to this fully merged result. The attention mechanism then caches the complete tensor $xW + xA_iB_i$ into a unified KV Cache. This design inherently binds the shared base projection $xW$ to adapter-specific updates, forcing every active agent to maintain an isolated, full-sized KV cache.

To eliminate this storage redundancy, \textsc{\SysName} introduces a disaggregated KV cache architecture, as depicted in Figure~\ref{fig:design-kv-cache}(b). For Key and Value processing, \textsc{\SysName} decouples the base model activations from the adapter states. The system first computes the base projection $xW$, applies RoPE to the Key tensors, and stores the resulting states in a unified base cache (\textit{bCache}). Concurrent requests with shared contexts can then access this \textit{bCache} via zero-copy sharing. To further minimize memory allocation, \textsc{\SysName} avoids computing the full adapter offset $xA_iB_i$. Instead, the system truncates the computation at the LoRA down-projection and stores the intermediate result $xA_i$ directly in a residual cache (\textit{rCache}). Notably, the RoPE operation is not applied to \textit{rCache} because their output dimensions mismatch. The low-rank property of the $A_i$ matrix ensures that the \textit{rCache} maintains a minimal memory footprint, drastically reducing the overall memory requirements of the serving system.

To quantify these memory savings mathematically, we formulate the memory consumption ratio $M_R$ between the unified and disaggregated KV cache architectures. Consider $N$ concurrent agents processing a shared context sequence of length $s$:

\begin{equation}
\begin{aligned}
    M_R = \frac{Mem_{disagg.}}{Mem_{unified}} &= \frac{Mem(xW) + N \cdot Mem(xA_i)}{N \cdot Mem(xW+xA_iB_i)} \\
    &= \frac{sn + N \cdot sr}{N \cdot sn} = \frac{1}{N} + \frac{r}{n}
\end{aligned}
\end{equation}

\noindent where $n$ denotes the output dimension of the base weight matrix $W$, and $r$ represents the rank of the LoRA down-projection matrix $A_i$. In standard configurations, $r \ll n$ (e.g., $n=1024$ and $r=16$). Consequently, as the number of concurrent agents $N$ grows large, the term $1/N$ approaches $0$, reducing the memory ratio to $M_R \approx r/n$. This theoretical bound demonstrates that \textsc{\SysName} drastically minimizes the per-agent memory consumption, allowing the system to support massive multi-LoRA agent workflows without memory exhaustion.

\subsection{Tree-Structured Cache with Fork Semantics}

\begin{figure}[t]
    \centering
    \includegraphics[width=\linewidth]{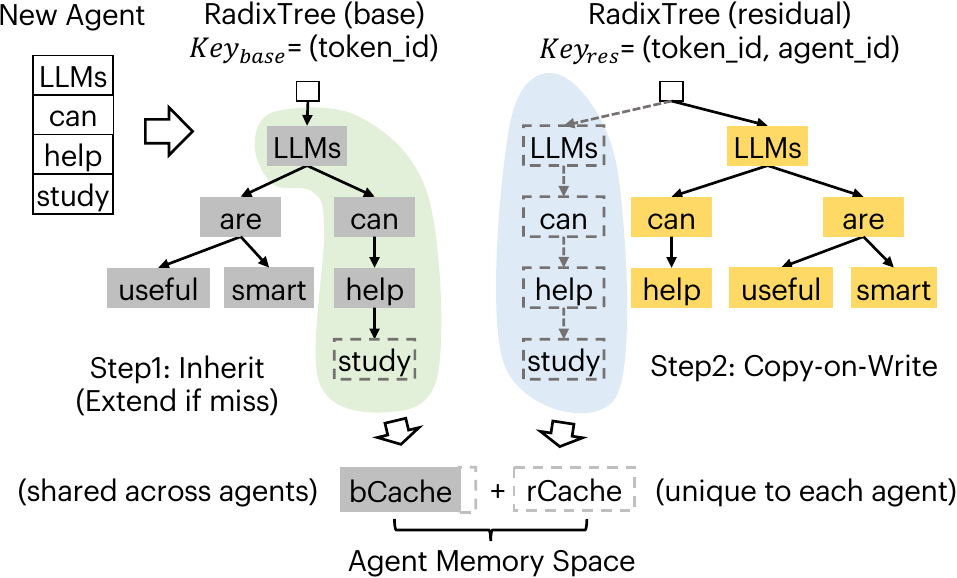}
    \caption{DualRadixTree Architecture. A new agent logically forks its memory space from former agents by inheriting the shared \textit{bCache} (Step1: Prefix Match) and exclusively allocating the \textit{rCache} (Step2: Copy-on-Write).}
    \label{fig:design-cache-tree}
\end{figure}

\noindent \textbf{DualRadixTree Architecture.} Managing a disaggregated KV cache requires a novel memory abstraction because traditional unified pools fail to accommodate the distinct sizes and lifecycles of shared and unique memory components. To bridge this gap and efficiently manage the memory space of concurrent agents, we introduce the DualRadixTree architecture that physically decouples the unified KV cache into a globally shared \textit{bCache} and an agent-specific \textit{rCache}, as shown in Figure~\ref{fig:design-cache-tree}. The \textit{bCache} is managed by a base RadixTree where the search key $Key_{base}$ is defined strictly by the sequence of token ids. This design guarantees that all concurrent requests possessing shared contexts can access the same underlying memory in the \textit{bCache} via zero-copy mechanisms. In parallel with the base structure, we deploy a residual RadixTree to index the \textit{rCache} and manage the distinct generation branches of individual agents. Because the \textit{rCache} is unique to each agent, the search key $Key_{res}$ in this residual tree extends the traditional token sequence with a specific agent id. Through this decoupled structural design, the DualRadixTree provides the exact memory abstraction required to materialize the disaggregated KV cache paradigm. By maintaining a unified logical memory view for each agent, this data structure successfully eliminates the need to allocate a full-sized KV cache for every agent as traditional prefix caching requires.

\noindent \textbf{Fork Semantics with CoW.} To orchestrate this decoupled memory architecture for incoming agent requests, \textsc{\SysName} introduces an OS-inspired fork semantics with CoW, as demonstrated by Figure~\ref{fig:design-cache-tree}. When a new agent is launched, the system first queries the base RadixTree to perform prefix matching to identify the longest shared context. If input tokens miss the existing prefix, the system dynamically extends the tree by allocating new shared memory blocks. The agent then inherits this context by mapping the globally shared \textit{bCache} into its logical memory space. This process is analogous to a newly forked OS process mapping the read-only physical pages of its parent. Following this inheritance, the system executes a CoW operation to allocate exclusive memory blocks for the \textit{rCache}. These unique allocations are tracked within the residual RadixTree to store an isolated state for the agent, functioning analogously to the private CoW pages of a child process. By orchestrating this explicit two-step allocation, our fork semantics establish a dynamic management mechanism for the logical memory space of agents, translating theoretical memory decoupling into a practical system design alongside the DualRadixTree architecture.

\noindent\textbf{Decoupled Eviction Policy.} In traditional prefix caching, the monolithic memory layout enforces a rigid eviction process: a cached sequence is evicted as an indivisible unit, resulting in a binary state of either a complete cache hit or a complete miss. However, directly applying this eviction policy to the disaggregated KV cache creates a strict architectural mismatch because the bCache and rCache exhibit fundamentally different memory footprints and access frequencies. A cascading eviction mechanism would couple these two memory pools, forcing a low-contention pool to discard active cache, thereby triggering entirely avoidable recomputation.

To eliminate this redundancy, \textsc{\SysName} introduces a decoupled eviction policy that isolates the lifecycle management of the base and residual caches by assigning independent Least Recently Used (LRU) states to each radix tree. This architectural isolation provides critical flexibility under heavy memory pressure. If a massive \textit{bCache} node is evicted while its lightweight \textit{rCache} counterpart persists, the scheduler avoids treating subsequent requests as complete cache misses. Instead, the system executes a \textit{partial hit} where the execution engine recomputes only the missing base projection $xW$, reinserts it into the base tree, and directly reuses the surviving $xA_i$ from the residual tree. This graceful degradation maximizes the overall cache hit rate for complex multi-round agent workflows.

\subsection{ResidualAttention}

Computing attention scores accurately requires reconstructing the KV cache from its disaggregated components. To minimize HBM memory and computational overhead, we fuse KV cache reconstruction directly into the attention computation within the SRAM. Driven by this architectural choice, we implement \textit{ResidualAttention}, an attention kernel specifically tailored for the disaggregated KV cache architecture. As illustrated in Algorithm~\ref{alg:res-attn}, our approach executes in three distinct stages. The first stage conducts on-the-fly Key cache reconstruction by streaming block-wise tiles and applying deferred RoPE operations. Next, the kernel computes attention scores independently for the base and residual components. Finally, the third phase leverages matrix associativity to fuse the attention output. By executing these steps entirely on-chip, \textit{ResidualAttention} maps the decoupled cache layout into a high-throughput parallel execution model. The following subsections detail the core algorithm and architectural innovations driving this design.

\begin{figure}[t]
    \centering
    \includegraphics[width=\linewidth]{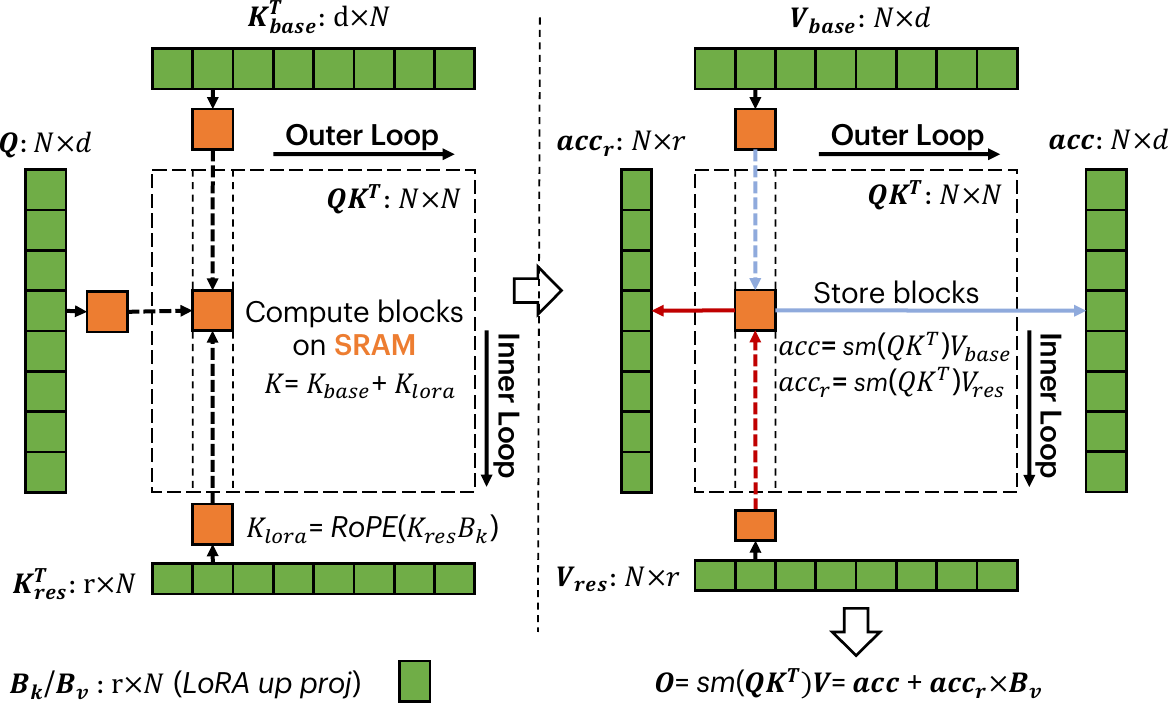}
    \caption{Illustration of ResidualAttention.}
    \label{fig:design-res-attn}
\end{figure}

\noindent \textbf{On-the-fly Key Reconstruction with Deferred RoPE.} The left portion of Figure~\ref{fig:design-res-attn} illustrates the cache reconstruction and attention computation process during the first stage. Our kernel streams the decoupled base cache $K_{base}$ and residual cache $K_{res}$ directly into the fast on-chip SRAM in a block-wise manner. Once inside the SRAM, the kernel reconstructs the complete cache on the fly. This reconstruction proceeds in two steps. First, we reconstruct the full-size LoRA Key cache $K_{lora}$ derived from $xA_iB_i$. We perform a LoRA up-projection using $K_{res}B_k$ and then apply RoPE to the resulting intermediate state. As detailed in Section~\ref{sec:design-kv-cache}, RoPE on \textit{rCache} is deferred during the earlier linear projection phase due to an output dimension mismatch. Applying it at this stage equips $K_{lora}$ with the exact token position information. Second, we materialize the combined Key cache using $K_{base}$ and the processed residual components $K_{lora}$. The kernel then computes the attention logits $QK^T$ and performs subsequent operations like the online softmax update. This integrated approach successfully avoids memory-expensive cache reconstruction operation in HBM and inefficient sequential in-place cache updates. Simultaneously, it guarantees the accuracy of the disaggregated KV cache transformation by ensuring the correct preservation of positional encodings.

\noindent \textbf{Fusing Attention Scores via Matrix Associativity.} After obtaining the attention logits, a straightforward approach to compute the final output $\sum sm(QK^T)V$ is to reconstruct the Value cache $V_{base}+V_{res}B_v$ inside the inner loop responsible for iterating over sequence blocks. This approach mirrors the reconstruction mechanism used for the Key cache. However, this eager projection method introduces prohibitive computational overhead and memory footprint. Executing the LoRA up-projection operation $V_{res}B_v$ at every loop step not only introduces significant redundant computation, but also demands large SRAM capacity to store this intermediate tensor, which inevitably triggers SRAM contention and severely degrades GPU parallelism.

\begin{algorithm}[t]
\caption{ResidualAttention}
\label{alg:res-attn}
\begin{algorithmic}[1]
\REQUIRE Query blocks $Q$
\REQUIRE \textit{bCache} blocks $(K_{base}, V_{base})$, \textit{rCache} blocks $(K_{res}, V_{res})$
\REQUIRE LoRA up-projection weights $B_k, B_v$
\STATE \textbf{Initialize:} $\text{acc} \leftarrow \mathbf{0}_{M \times D_v}$, $\text{acc}_r \leftarrow \mathbf{0}_{M \times R}$
\STATE \textbf{Initialize states:} $m \leftarrow -\infty$, $l \leftarrow 0$
\STATE Load $B_k$, $B_v$ to SRAM
\FOR{each block $n$ in sequence}
    \STATE \algcomment{Stage1: On-the-fly Key Reconstruction with Deferred RoPE}
    \STATE Load $Q$, $K_{base}$, $V_{base}$, $K_{res}$, $V_{res}$ to SRAM
    \STATE Load cached sin-cos table for RoPE to SRAM
    \STATE $K_{lora} \leftarrow \text{RoPE}(K_{res} \cdot B_{k})$
    \STATE $K \leftarrow K_{base} + K_{lora}$
    \STATE \algcomment{Stage2: Compute Separate Attention Scores (base/residual)}
    \STATE $S \leftarrow Q \cdot K^T \cdot \text{scale}$
    \STATE $m_{new} \leftarrow \max(m, \max(S))$
    \STATE $l_{new} \leftarrow l \cdot \exp(m - m_{new}) + \sum \exp(S - m_{new})$
    \STATE $P \leftarrow \exp(S - m_{new})$
    
    \STATE $\text{acc} \leftarrow \text{acc} \cdot \exp(m - m_{new}) + P \cdot V_{base}$
    \STATE $\text{acc}_r \leftarrow \text{acc}_r \cdot \exp(m - m_{new}) + P \cdot V_{res}$
    
    \STATE $m \leftarrow m_{new}$, $l \leftarrow l_{new}$
\ENDFOR
\STATE \algcomment{Stage3: Fuse Attention Scores via Matrix Associativity}
\STATE $\text{acc}_{final} \leftarrow \text{acc} + \text{acc}_r \cdot B_v$
\STATE Write $O \leftarrow \text{acc}_{final} / l$ to HBM
\RETURN $O$
\end{algorithmic}
\end{algorithm}

To resolve these inefficiencies, we decouple the attention computation for the base and residual components, and then fuse their partial outputs at the very end of the kernel execution. The right portion of Figure~\ref{fig:design-res-attn} illustrates both this decoupled computation and delayed projection process. This strict mathematical equivalence is guaranteed by the associativity of matrix multiplication:

\begin{equation}
\begin{aligned}
    \sum sm(QK^T)V &= \sum sm(QK^T)\cdot(V_{base}+V_{res}B_v) \\
                     &= \sum sm(QK^T)V_{base} + (\sum sm(QK^T)V_{res})B_v
\end{aligned}
\end{equation}

\noindent where $V_{base}$ and $V_{res}$ denote the base and residual Value cache respectively, and $B_v$ is the LoRA up-projection weight matrix. We push the $B_v$ multiplication entirely out of the inner loop and compute this up-projection only once at the end of the kernel. To support the decoupled attention computation, the kernel maintains a lightweight global accumulator for residual attention scores alongside the base cache. By leveraging this approach, we drastically minimize both computational overhead and the SRAM allocation.

\section{Implementation}

We implement \SysName on top of SGLang (v0.5.6) with approximately 3K lines of Python code and custom Triton kernels. Our implementation introduces system-wide modifications across the control plane, model executor, and kernel layers to natively support the disaggregated KV cache layout.

\noindent\textbf{Disaggregated KV Cache.} To support the physical decoupling of the KV cache at runtime, we design a custom LoRA replacement module for the linear projection layer. This module separates the adapter-specific residual activations from the base model activations and store these residual activations into a dedicated KV cache pool indexed by the residual RadixTree.

\noindent\textbf{Control Plane and DualRadixTree Storage.} We extend the native RadixCache of SGLang into a coordinated DualRadixTree architecture to manage the decoupled memory layout. We also adapt the scheduler to orchestrate this two-tiered KV cache pool across phases like chunked prefill, non-chunked prefill, and decode.

\noindent\textbf{ResidualAttention.} We implement this custom hardware-aware attention kernel in Triton. This kernel is adapted from the RadixAttention kernel of SGLang and have two separate versions for prefill and decode. This structural separation allows the attention mechanism to cater specifically to the distinct memory access patterns and workload characteristics of each execution phase.

\section{Evaluation}

\subsection{Setup}

\noindent \textbf{Models and Hardware Settings.} We evaluate \textsc{\SysName} on three open-source large language models: Llama3-8B~\cite{llama3modelcard}, Qwen2.5-7B~\cite{qwen2.5}, and Qwen2.5-14B~\cite{qwen2.5}. All models are deployed in BF16 precision. We conduct our end-to-end experiments across two hardware platforms: a server equipped with a single L40 GPU and 128 vCPUs, and a server containing two RTX 5000 GPUs with 48 vCPUs. Specifically, Llama3-8B is served on the L40 GPU, while Qwen2.5-7B and Qwen2.5-14B are evaluated on one and two RTX 5000 GPUs respectively.

\vspace{0.5em}

\noindent \textbf{System Performance Benchmark.} To evaluate the serving throughput and efficiency of \textsc{\SysName}, we benchmark under two representative agentic execution patterns:

\begin{itemize}[leftmargin=*]
    \item \textit{ReAct~\cite{yao2022react}:} An iterative reasoning and acting paradigm where the context window grows sequentially. We use it to evaluate KV cache management efficiency during sustained, multi-turn agent workloads.
    \item \textit{MapReduce~\cite{luo2025autellix}:} A parallel processing workflow that divides a large task into multiple concurrent subtasks. We use it to stress-test \textsc{\SysName}'s memory sharing capability when handling massive simultaneous forks from a single shared context.
\end{itemize}

\begin{figure*}[t]
    \centering
    \includegraphics[width=0.75\linewidth]{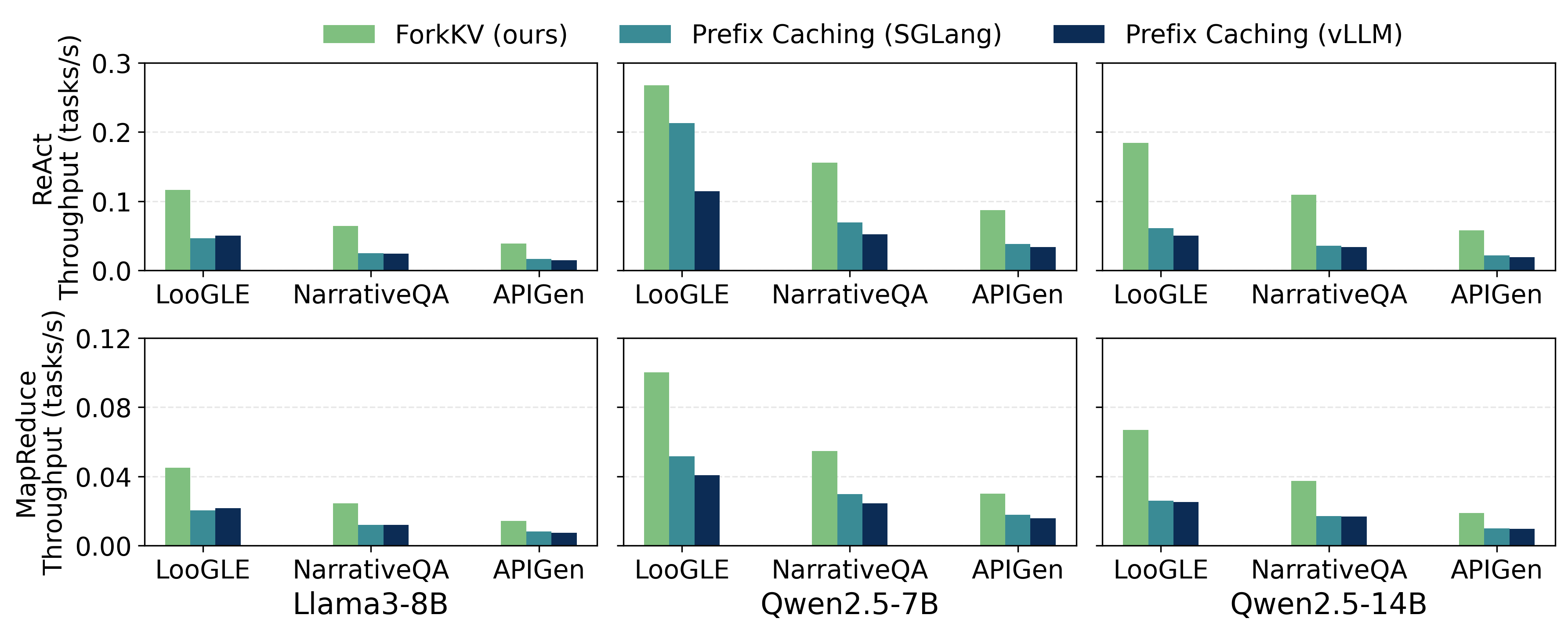}
    \caption{End-to-end throughput evaluation. The figure compares the serving throughput (tasks/s) of \textsc{\SysName} against prefix caching baselines on ReAct and MapReduce workloads across three models and three datasets.}
    \label{fig:e2e}
\end{figure*}

\noindent To simulate the scenarios where concurrent agents operate on the same context and execute different tasks, we synthesize the model input using two components: a massive static part shared across all agents and a dynamic part containing task-specific instructions. We construct these two-stage inputs using three long-context datasets, as summarized in Table~\ref{tab:dataset}. \textit{LooGLE}~\cite{li2023loogle} features long documents from various sources such as arXiv, Wikipedia, and movie/TV scripts. In our evaluation, we leverage the movie/TV scripts as the static part and use the corresponding questions as the dynamic part. \textit{NarrativeQA}~\cite{kocisky2018narrativeqa} is a widely acknowledged long-context dataset for testing reading comprehension capabilities, featuring documents even longer than those in LooGLE. We follow a similar methodology here by using the narrative document as the static part and the comprehension question as the dynamic part. Finally, \textit{APIGen}~\cite{liu2024apigen} is a tool-calling dataset designed to evaluate instruction-following and structured output capabilities. We aggregate the available tool descriptions to construct the large shared static context and use the specific instructions as the dynamic inputs.

Furthermore, we configure 8 agent workflows for both the ReAct and MapReduce paradigms, where baselines face severe performance degradation. In these workflows, each individual agent utilizes a distinct LoRA adapter with a rank of 16, following the setting in prior works~\cite{zhu2025cannikin,chen2024punica}. Figure~\ref{fig:motivation-context-reuse} illustrates the underlying architecture of these workflows. We generate continuous requests with an average arrival rate of 2 requests per second to evaluate the system under sustained heavy load. Within each workflow loop, we simulate agent-tool interactions by injecting a constant latency of 0.1 seconds and returning a mock tool response of 100 random tokens. For the agent generation step, we set the maximum output length to 256 tokens.

\vspace{0.5em}

\begin{table}[t]
\centering
\begin{tabular}{cccc}
\toprule
& LooGLE & NarrativeQA & APIGen \\
\midrule
Static Context & 32742 & 49119 & 64911 \\
Avg. Dynamic Instr & 24 & 12 & 23 \\
\bottomrule
\end{tabular}
\caption{The length of shared static context and task-specific dynamic instructions (average) in the sampled datasets.}
\label{tab:dataset}
\end{table}

\noindent \textbf{System Performance Baselines.} We compare \textsc{\SysName} against two baselines in the system performance benchmark:

\begin{itemize}[leftmargin=*]
    \item \textit{vLLM~\cite{kwon2023efficient}:} A state-of-the-art serving engine that utilizes PagedAttention for efficient KV cache memory management. It also implements prefix caching for efficient KV cache reuse. We use vLLM v0.12.0 as our baseline.
    \item \textit{SGLang~\cite{zheng2024sglang}:} Another state-of-the-art framework featuring RadixAttention, which enables automatic KV cache reuse for shared-prefix requests. We use SGLang v0.5.6 as our baseline.
\end{itemize}

\noindent We select SGLang and vLLM as our baselines because they provide state-of-the-art LLM serving capabilities. Both frameworks integrate a comprehensive set of performance acceleration features, ranging from widely adopted mechanisms like prefix caching to advanced techniques such as CUDA Graphs and asynchronous scheduling. While dedicated LoRA serving systems~\cite{sheng2023slora,chen2024punica,wu2024dlora} also provide specialized LoRA serving capabilities, they generally lack full support for these system-level optimizations.

\vspace{0.5em}

\begin{figure}[t]
    \centering
    \includegraphics[width=\linewidth]{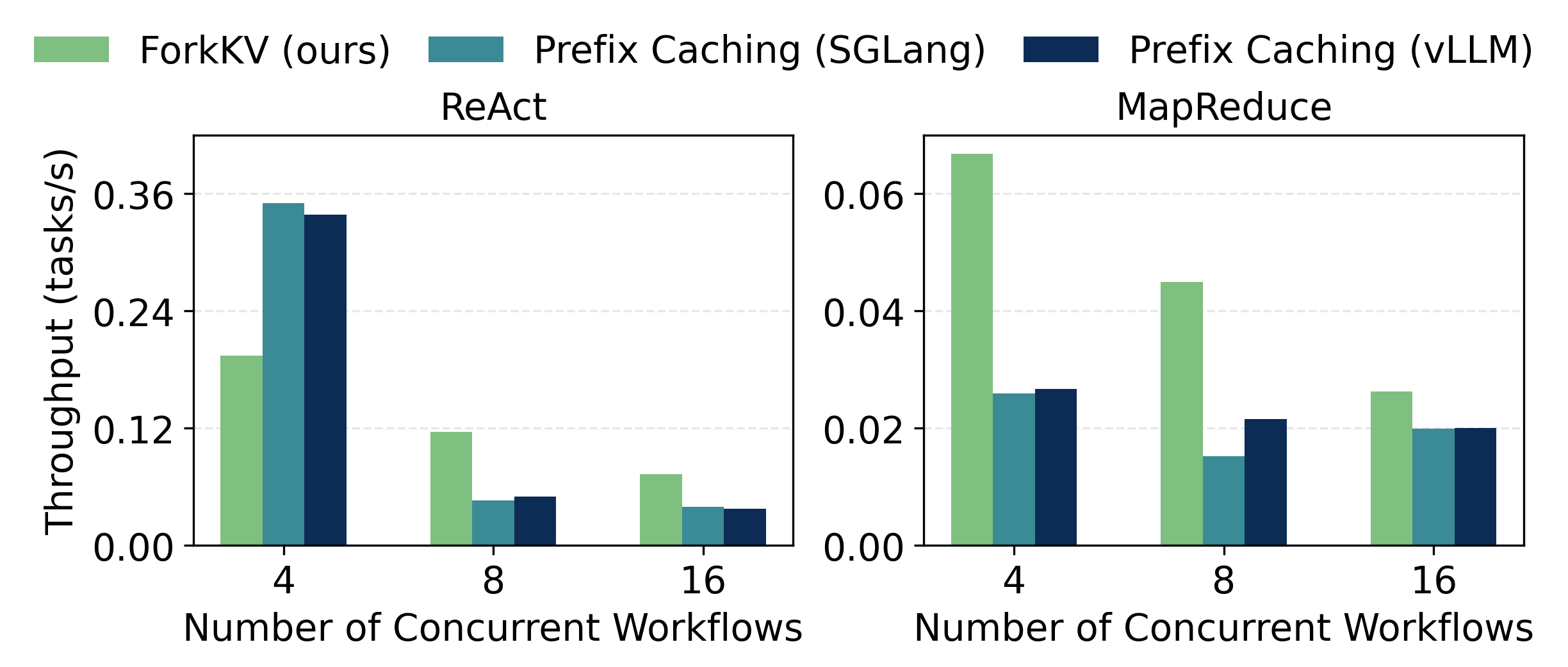}
    \caption{Performance under varying number of workflows with Llama3-8B on the LooGLE dataset. Every individual agent in different workflows uses different LoRA adapters.}
    \label{fig:e2e-parallelism}
\end{figure}

\noindent \textbf{Generation Quality Benchmark.} To verify that \textsc{\SysName} introduces negligible accuracy loss, we evaluate the generation quality using the following datasets and metrics:

\begin{itemize}[leftmargin=*]
    \item \textit{HotpotQA}~\cite{yang2018hotpotqa}: This is a multi-hop question answering dataset. It's intended to evaluate the model's complex reasoning and information integration capabilities across multiple contexts. We include 200 test cases.
    \item \textit{APIGen}~\cite{liu2024apigen}: In addition to system benchmarking, we utilize 200 test cases from APIGen to benchmark the model's instruction-following and structured API calls generation capabilities.
\end{itemize}

\noindent We adopt \textit{F1-score}~\cite{ffl2020evalqa}, a metric that measures the similarity between the model's output and the ground-truth answer of the question based on the number of overlapping words.

\vspace{0.5em}

\noindent \textbf{Generation Quality Baselines.} We compare \textsc{\SysName} against two KV cache sharing policies in the generation quality benchmark:

\begin{itemize}[leftmargin=*]
    \item \textit{Prefix Caching}~\cite{kwon2023efficient,zheng2024sglang}: This policy strictly reuses the KV cache of an identical context only when the requests share the exact same LoRA adapter. It is mathematically lossless and serves as the accuracy upper bound for our evaluation.
    \item \textit{Full Reuse}: This policy aggressively reuses the KV cache of a shared context across different LoRA adapters, entirely ignoring the distinct activations introduced by each adapter. We include this to establish a baseline for the accuracy drop that \textsc{\SysName} aims to mitigate.
\end{itemize}

\begin{figure}[t]
    \centering
    \includegraphics[width=\linewidth]{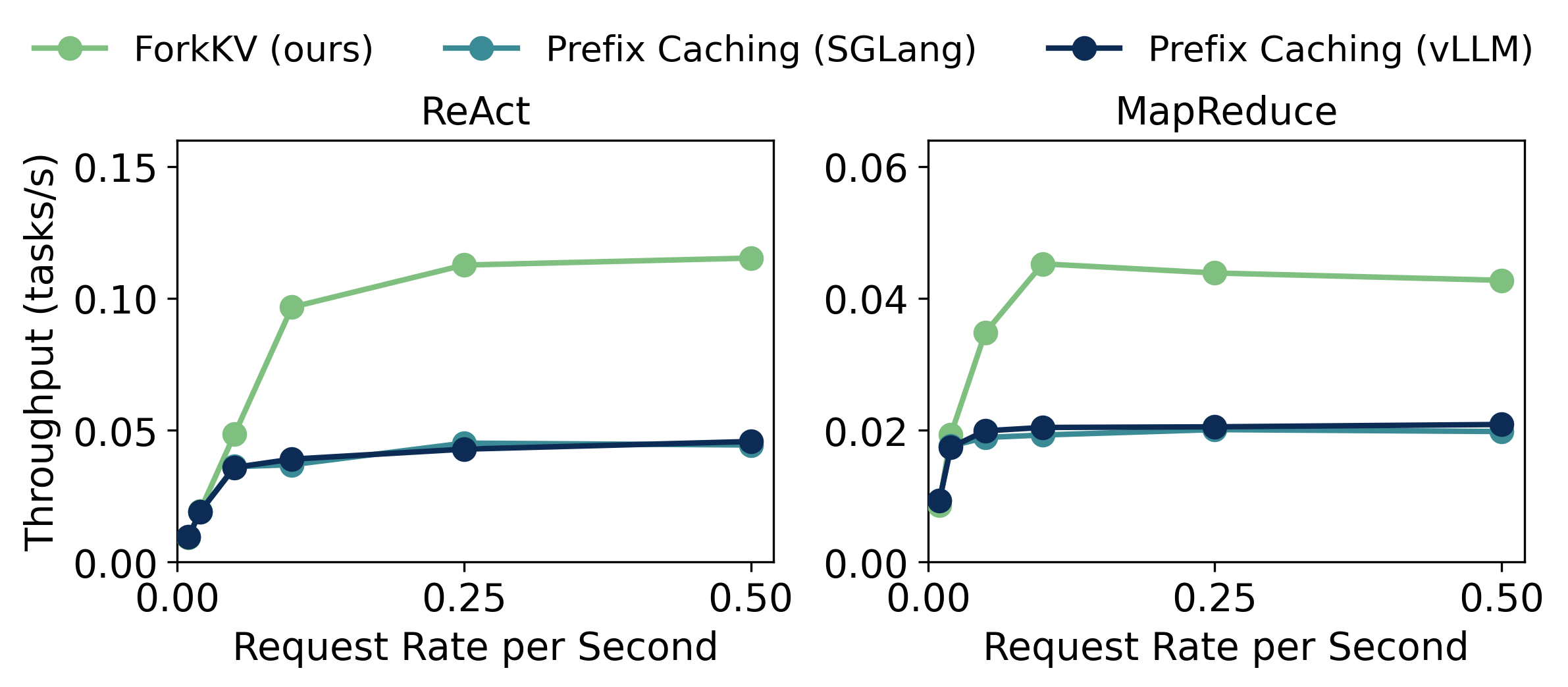}
    \caption{Performance under different requests arrival rate with Llama3-8B on the LooGLE dataset.}
    \label{fig:e2e-qps}
\end{figure}

\subsection{End-to-End Results}

\noindent \textbf{Throughput Improvement.} Figure~\ref{fig:e2e} compares the end-to-end throughput of \textsc{\SysName} against prefix caching baselines across three datasets and three models under the ReAct and MapReduce paradigms. Overall, \textsc{\SysName} achieves 1.25$\times$ to 3.04$\times$ the throughput of the baselines on ReAct workflows, and 1.68$\times$ to 2.60$\times$ the throughput on MapReduce workflows. These consistent speedups demonstrate the scalability of \textsc{\SysName} in handling both sequential and concurrent agent workflows. Beyond the overall gains, we draw several key insights from different system configurations. First, \textsc{\SysName} yields the most significant speedups in scenarios with severe memory contention. For instance, \textsc{\SysName} achieves 3.04$\times$ the baseline throughput on the larger Qwen2.5-14B model but only 1.25$\times$ on the smaller Qwen2.5-7B model using the LooGLE dataset under ReAct workflow. The smaller model requires less memory for base weights and the KV cache. This reduced memory footprint makes the memory savings from \textsc{\SysName} less prominent while amplifying the computational overhead of the disaggregated KV cache design. Second, \textsc{\SysName} maintains robust performance gains across datasets with varying sequence lengths. Although the throughput values vastly differ across datasets, the relative speedups on LooGLE, NarrativeQA, and APIGen remain stable at 1.20$\times$, 1.32$\times$, 1.09$\times$ respectively. This stability confirms that \textsc{\SysName} remain highly effective regardless of task type or sequence length.

\noindent \textbf{Performance under Varying Number of Workflows.} Figure~\ref{fig:e2e-parallelism} illustrates the throughput of \textsc{\SysName} compared to prefix caching baselines as the number of concurrent agent workflows scales. \textsc{\SysName} initially exhibits lower throughput than the baselines under a light load of 4 ReAct workflows. At this low concurrency level, the KV cache consumption of active agents remains small relative to the available GPU memory. Standard prefix caching can therefore retain all KV cache blocks without triggering evictions, preserving high baseline performance by avoiding costly recomputations. In contrast, \textsc{\SysName} introduces disaggregated KV cache with dedicated architectural and kernel support to reduce the per-agent memory footprint. However, this specialized design incurs noticeable computational overhead when GPU memory is abundant, which accounts for the initial performance drop. This limitation can be mitigated by adaptive scheduling, which monitors GPU memory utilization and dynamically falls back to standard KV cache when memory is abundant. Despite the initial performance gap, the architectural design of \textsc{\SysName} yields significant benefits as the number of concurrent workflows scales and memory demand exceeds hardware capacity. Under these high-contention conditions, \textsc{\SysName} achieves 1.84-2.33$\times$ and 1.31-2.51$\times$ the throughput of the baselines on ReAct and MapReduce paradigms respectively. These results confirm that \textsc{\SysName} successfully delivers high performance during heavy memory contention.

\begin{figure}[t]
    \centering
    \includegraphics[width=\linewidth]{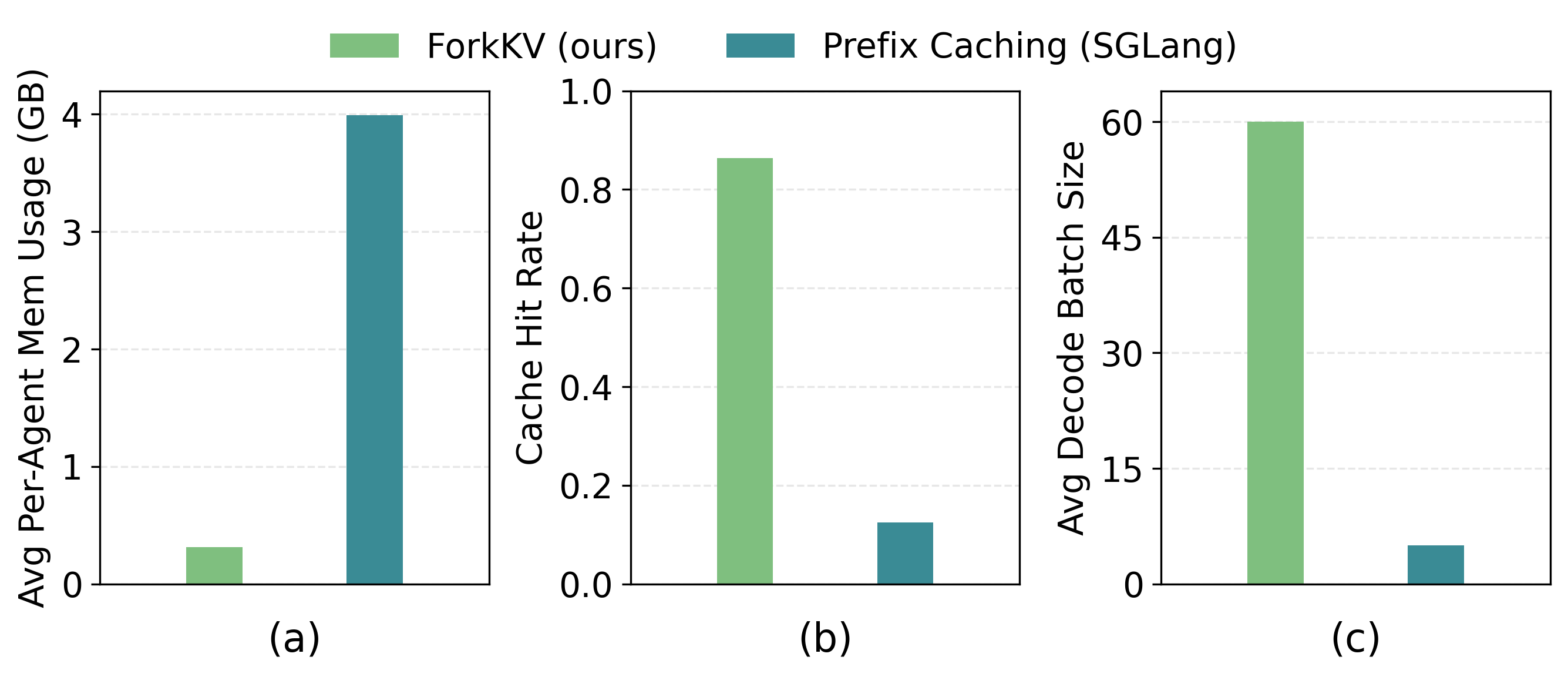}
    \caption{Underlying causes of \textsc{\SysName} performance gains. \textsc{\SysName} reduces (a) the average per-agent memory usage, which consequently improves (b) the cache hit rate and increases (c) the average decode batch size.}
    \label{fig:e2e-principle}
\end{figure}

\noindent \textbf{Performance under Different Requests Arrival Rate.} Figure~\ref{fig:e2e-qps} demonstrates the throughput of \textsc{\SysName} compared to prefix caching baselines as the request arrival rate increases. \textsc{\SysName} consistently outperforms the baselines across varying request rates. As the arrival rate scales, standard prefix caching struggles with costly recomputation caused by frequent cache evictions. Conversely, \textsc{\SysName} manages this increased load efficiently due to the shrinked per-agent memory footprint enabled by disaggregated KV cache. Consequently, \textsc{\SysName} achieves approximately 2.52$\times$ and 2.05$\times$ the throughput of the baselines during the steady state. These results validate that \textsc{\SysName} successfully maintains high performance across a wide range of request arrival rates.

\noindent \textbf{Understanding \textsc{\SysName}'s Improvement.} \textsc{\SysName} achieves higher throughput by significantly reducing the per-agent memory footprint. Specifically, while prefix caching stores a separate KV cache for each LoRA-based agent, \textsc{\SysName} alleviates this inefficiency by decoupling the KV cache into a shared full-sized \textit{bCache} and a LoRA-specific lightweight \textit{rCache}. As Figure~\ref{fig:e2e-principle}a illustrates, this design reduces the average per-agent memory footprint by 12.7$\times$ compared to traditional prefix caching. This drastic reduction in memory consumption translates to performance gains in two distinct ways. First, the freed GPU capacity allows \textsc{\SysName} to retain more agent contexts within a limited GPU capacity. Consequently, as Figure~\ref{fig:e2e-principle}b shows, the cache hit rate of \textsc{\SysName} improves by 6.93$\times$ over prefix caching baselines, which directly accelerates generation by reducing the frequency of recomputation. Second, the lower memory footprint per agent enables a 12.0$\times$ larger decode batch size as demonstrated in Figure~\ref{fig:e2e-principle}(c). This expanded batch size allows for greater parallelism during agent serving. In practice, these two advantages hide the overhead introduced by the new architecture and deliver substantial speedups.

\begin{figure}[t]
    \centering
    \includegraphics[width=0.9\linewidth]{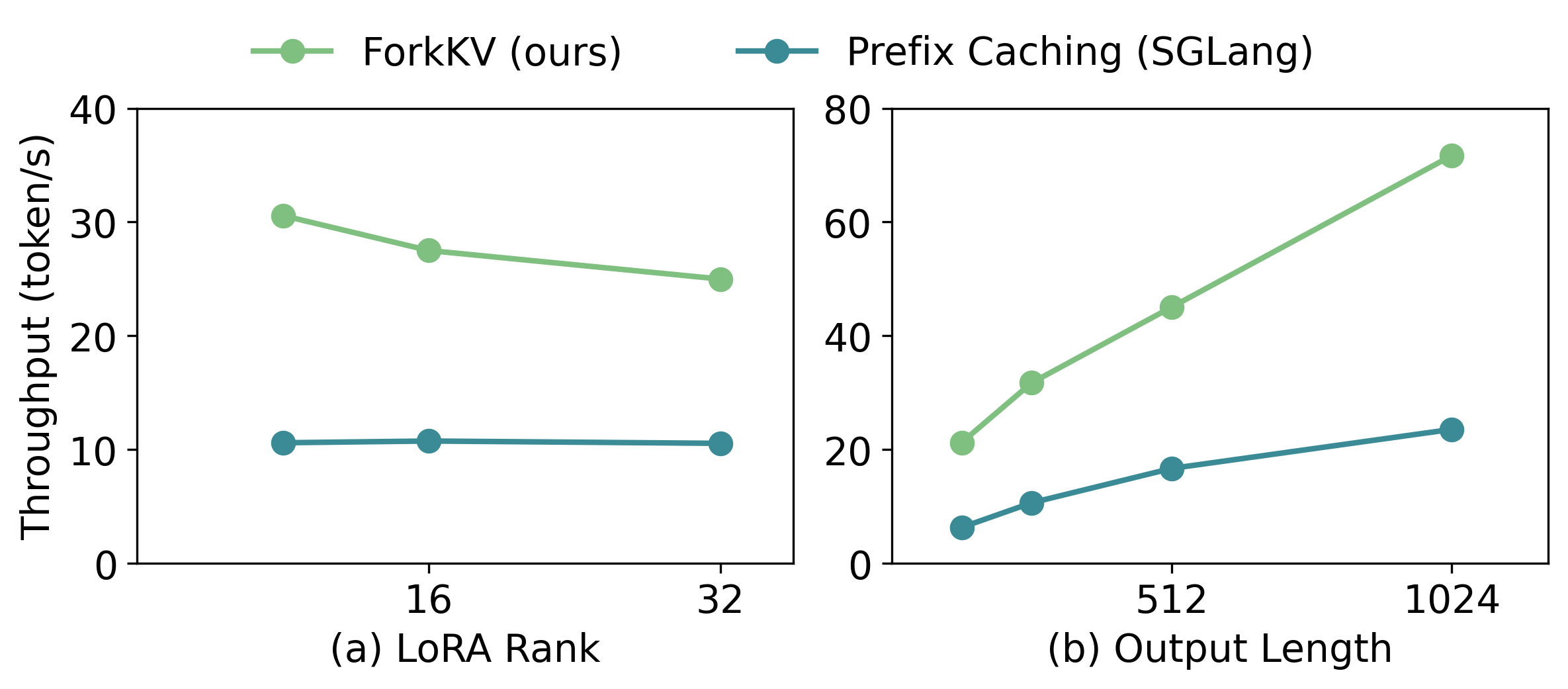}
    \caption{Sensitivity study of varying LoRA ranks and output lengths with Llama3-8B on the ReAct workflow.}
    \label{fig:sensitivity}
\end{figure}

\begin{table}[t]
\centering
\begin{tabular}{cccc}
\toprule
Model & Sharing Policy & HotpotQA & APIGen \\
\midrule
& Prefix Caching & 57.63 & 39.77 \\
Llama3-8B & \SysName & 57.17 & 38.17 \\
& Full Reuse & 54.02 & 17.82 \\
\midrule
& Prefix Caching & 57.14 & 92.28 \\
Qwen2.5-7B & \SysName & 56.37 & 91.52 \\
& Full Reuse & 55.47 & 90.08 \\
\midrule
& Prefix Caching & 70.91 & 94.56 \\
Qwen2.5-14B & \SysName & 70.66 & 94.16 \\
& Full Reuse & 68.86 & 93.66 \\
\bottomrule
\end{tabular}
\caption{Generation quality evaluated with F1-score (\%).}
\label{tab:accuracy}
\end{table}

\subsection{Accuracy Verification}

Section~\ref{sec:motivation-sharing} suggests that \textsc{\SysName} introduces negligible degradation in generation quality. To verify this claim, we evaluate \textsc{\SysName} against standard prefix caching and full KV cache reuse across three distinct models and two datasets. Table~\ref{tab:accuracy} summarizes these comparative results. The evaluation shows that \textsc{\SysName} achieves an accuracy highly comparable to the prefix caching baseline, exhibiting an average quality drop of only 0.71 points across all evaluated settings. Furthermore, the maximum observed decrease is just 1.60 points on Llama3-8B when tested with the APIGen dataset. In contrast, full KV cache reuse suffers from severe performance degradation and incurs an average accuracy drop of 5.40 points. This degradation is particularly obvious on complex tasks like APIGen, where the accuracy of Llama3-8B drops by a substantial 21.95 points from 39.77 to 17.82. By avoiding these extreme performance penalties, \textsc{\SysName} successfully preserves generation quality across diverse models and tasks while enabling efficient KV cache sharing.

\subsection{Sensitivity Analysis}

To have a better understanding of \textsc{\SysName}, we conduct sensitivity analysis across different configurations with Llama3-8B on the ReAct workflow.

\noindent \textbf{Varying LoRA Ranks.} To study the impact of LoRA ranks on \textsc{\SysName}, we modify the underlying LoRA ranks of agents across a representative set $r\in \{8, 16, 32\}$~\cite{chen2024punica,sheng2023slora,wu2024dlora,zhu2025cannikin}. Figure~\ref{fig:sensitivity}a shows \textsc{\SysName} achieves 2.36-2.88$\times$ the baseline throughput over this setting. Noticeably, the absolute throughput of \textsc{\SysName} decreases as the rank increases because a larger rank linearly expands the residual cache size, which increases the memory footprint per agent and limits the maximum batch size during inference. However, since small ranks ($r < 64$) can already provide high generation quality for various NLP tasks~\cite{hu2022lora}, \textsc{\SysName} is highly effective in serving multi-LoRA agents under practical configurations.

\noindent \textbf{Varying Output Lengths.} We evaluate the system by varying the output length of each agent within the workflow. Figure~\ref{fig:sensitivity}b shows \textsc{\SysName} achieves 2.69-3.36$\times$ the baseline throughput across different lengths. Longer agent outputs continuously accumulate newly generated KV cache and severely challenge the system memory capacity. However, as discussed in Section~\ref{sec:motivation-sharing}, \textsc{\SysName} consistently maintains a low memory footprint per agent. This minimal memory overhead allows our system to absorb the growing memory demands and sustain a strictly larger concurrent batch size than the baseline. Consequently, \textsc{\SysName} guarantees highly efficient serving performance regardless of the specific generation length.

\section{Related Work}

\noindent\textbf{Agentic Workflow Serving.} The rapid evolution of autonomous agents has spurred the development of specialized serving systems for complex agent workflows, which mainly optimize via efficient scheduling~\cite{luo2025autellix,dai2025aragog,lin2024parrot,fu2024certaindex,liu2025circinus,chaudhry2025murakkab} and KV cache routing~\cite{pan2025kvflow,bian2025tokencake,wu2026dualpath} for general-purpose LLMs. In contrast, ForkKV targets multi-LoRA agent serving, enabling base cache sharing across distinct agents to minimize memory footprint and alleviate batch size restrictions.

\noindent\textbf{System-level Optimization on LoRA Systems.} The popularity of LoRA has driven the need for efficient multi-tenant LoRA systems where numerous adapters share a single base model~\cite{chen2024punica,sheng2023slora,nikoleta2024chameleon,wu2024dlora,li2025toppings,zhang2025fastlibra,chen2025mixlora,zhou2024dynamicoperator,zhu2025cannikin,zhu2025lorafusion,ye2025mlora}. Pioneering works like Punica~\cite{chen2024punica} introduce custom CUDA kernels to efficiently batch requests across different adapters. Building upon this foundation, subsequent research has explored advanced scheduling policies~\cite{wu2024dlora,chen2025mixlora,zhu2025cannikin}, memory optimizations~\cite{sheng2023slora,nikoleta2024chameleon}, GPU kernel optimizations~\cite{zhou2024dynamicoperator,zhu2025lorafusion,ye2025mlora}, and offloading techniques~\cite{li2025toppings,zhang2025fastlibra} for efficient serving. While prior works primarily target chatbot scenarios, ForkKV tackles the severe memory redundancy caused by diverging LoRA activcations in multi-agent workflows through KV cache disaggregation.

\noindent\textbf{Copy-on-Write (CoW) in Data Management Systems.} Originating as a classic OS-level technique to enable lock-free parallel operations and reduce memory footprints, CoW is widely used to optimize data management systems~\cite{kemper2011hyper,wang2023dlsm,lakshman2022magma,okolnychyi2024petabyte,cha2023blink,fruth2024case}. Inspired by these approaches, ForkKV adapts the CoW mechanism to manage the KV cache for highly-branched shared contexts across agents, effectively extending this paradigm to multi-LoRA agent serving scenarios.

\noindent\textbf{KV Cache Optimization.} Existing KV cache optimization strategies primarily focus on lossless memory layout improvements~\cite{zheng2024sglang,xie2025strata,liu2025mell,xiong2024layerkv,qin2024mooncake}, lossy compression~\cite{liu2024cachegen,yao2025deltazip}, and cross-chunk or cross-model sharing~\cite{gim2024promptcache,yao2025cacheblend,liu2024droidspeak}. In concurrent work, LRAgent~\cite{jeon2026lragent} proposes to decompose the KV cache into shared and adapter-dependent components for multi-LoRA agent serving with negligible accuracy loss. Distinct from these approaches, our work uniquely introduces an OS-inspired DualRadixTree for decoupled cache management and an efficient attention kernel fused with deferred RoPE operations.


\section{Conclusion}

In this paper, we present \SysName, a highly efficient multi-LoRA agent serving system that resolves the critical memory footprint bottlenecks caused by KV cache divergence. Inspired by the OS fork primitive with copy-on-write, \SysName utilizes a DualRadixTree architecture to disaggregate the KV cache into a globally shareable base and lightweight adapter-specific residuals. To make this disaggregated memory layout computationally efficient, we design \textit{ResidualAttention}, an attention kernel that fuses cache reconstruction directly in SRAM. Our experiments demonstrate that \SysName achieves up to 3.0$\times$ throughput of state-of-the-art serving systems while preserving generation quality.

\begin{acks}
We would like to express our sincere gratitude to Zhanda Zhu from the University of Toronto for his valuable contributions to this research work. His expertise, dedication, and generous support have significantly enhanced the quality of our study. His insightful suggestions and technical assistance were instrumental in achieving our research objectives. We are also grateful to Zhihao Jia from Carnegie Mellon University for his valuable discussions during the early stages of this work. His perspectives and insights helped shape the foundation of our research.

\end{acks}

\bibliographystyle{ACM-Reference-Format}
\bibliography{reference}


\end{document}